\documentclass[aps,prb,twocolumn,english,showpacs,superscriptaddress,amssymb,amsfonts]{revtex4}

\usepackage[T1]{fontenc}
\usepackage[latin9]{inputenc}
\usepackage{amsmath}
\usepackage{dsfont}
\usepackage{amstext}
\usepackage{tocvsec2}

\usepackage{amssymb}
\usepackage{amsbsy}
\usepackage{amsthm}
\usepackage{epsfig}
\usepackage{graphicx}
\usepackage{bbm}
\usepackage[breaklinks]{hyperref}
\hypersetup{colorlinks=true,allcolors=blue}
\usepackage{color}
\usepackage{multirow}
\usepackage{pdfsync}
\usepackage{babel}

\begin{document}
	\title{Classification and properties of quantum spin liquids on the hyperhoneycomb lattice}
	\author{Biao Huang}
\affiliation{Department of Physics and Astronomy, University of Pittsburgh, Pittsburgh PA 15260, USA}
\author{Wonjune Choi}
\affiliation{Department of Physics and Centre for Quantum Materials,
		University of Toronto, Toronto, Ontario M5S 1A7, Canada}
\author{Yong Baek Kim}
\affiliation{Department of Physics and Centre for Quantum Materials,
		University of Toronto, Toronto, Ontario M5S 1A7, Canada}
\affiliation{Canadian Institute for Advanced Research/Quantum Materials Program, Toronto, Ontario MSG 1Z8, Canada}
\affiliation{School of Physics, Korea Institute for Advanced Study, Seoul 130-722, Korea}
\author{Yuan-Ming Lu}
	\affiliation{Department of Physics, The Ohio State University, Columbus, OH 43210, USA}
	\date{\today}
	\begin{abstract}
		\noindent The family of ``Kitaev materials'' provides an ideal platform to study quantum spin liquids and their neighboring magnetic orders. Motivated by the possibility of a quantum spin liquid ground state in pressurized hyperhoneycomb iridate $ \beta $-Li$ _2 $IrO$ _3 $, we systematically classify and study symmetric quantum spin liquids on the hyperhoneycomb lattice, using the Abrikosov-fermion representation. Among the 176 symmetric $U(1)$ spin liquids (and 160 $Z_2$ spin liquids), we identify 8 ``root'' $U(1)$ spin liquids in proximity to the ground state of the solvable Kitave model on hyperhonecyomb lattice. These 8 states are promising candidates for possible $U(1)$ spin liquid ground states in pressurized $ \beta $-Li$ _2 $IrO$ _3 $. We further discuss physical properties of these 8 $U(1)$ spin liquid candidates, and show that they all support nodal-line-shaped spinon Fermi surfaces.  
	\end{abstract}
	\maketitle

	\section{Introduction}

The exactly solvable Kitaev model on honeycomb lattice\cite{Kitaev2006} provides one pristine platform to study the physical properties of quantum spin liquids\cite{Savary2016,Zhou2017}, a class of long-range entangled many-body ground states featuring fractionalized excitations. Much attention is drawn to the spin liquid physics of Kitaev model, motivated by the proposal to design Kitaev exchange interactions in 4$d$ and 5$d$ transition-metal-based insulators with strong spin-orbit coupling\cite{Jackeli2009}. Following this proposal, a class of so-called ``Kitaev materials'' have been extensively studied both theoretically and experimentally\cite{Rau2016,Schaffer2016,Hermanns2017,Winter2017}, whose physics is believed to be in close proximity to the solvable Kitaev model.

In particular, $\beta$-Li$_2$IrO$_3$\cite{Biffin2014,Takayama2015} is a three-dimensional (3d) Kitaev material, where the low-energy $j_\text{eff}=\frac12$ magnetic moments on the iridium sites form a trivalent 3d network coined the ``hyperhoneycomb'' lattice (see FIG. \ref{fig:lattice}). The Kitaev model on the hyperhoneycomb lattice can be solved exactly and the quantum ground state is a 3d quantum spin liquid with nodal-line Majorana fermions~\cite{Schaffer2015,Halasz2017,OBrien2016,Kimchi2014}. Though this hyperhoneycomb iridate forms an incommensurate spiral magnetic order below $T_N\approx37$ K\cite{Biffin2014,Takayama2015}, recent X-ray magnetic circular dichroism measurements\cite{Veiga2017} showed that the zero-field magnetic order can be completely suppressed by applying hydrostatic pressure of 2 GPa\cite{Kim2016,Veiga2017}. Moreover, 
the paramagnetic state is maintained up to about 4 GPa, after which there is a structural phase transition. It has been known that the presence of the Kitaev interaction (as well
as other additional interactions) is crucial to explain the incommensurate spiral magnetic order at ambient pressure \cite{Lee2015, Lee2016, Ducatman2018}. Because of this, it has been suggested that the paramagnetic state between 2 GPa and 4 GPa could be a quantum spin liquid derived from the proximate Kitaev spin liquid \cite{Veiga2017, Kim2016}. 

In this work, motivated by the experimental progress mentioned above, we consider the possibility of a symmetric quantum spin liquid phase in pressurized $\beta$-Li$_2$IrO$_3$. We first provide the full classification of symmetric spin liquids on the hyperhoneycomb lattice within the fermionic parton (i.e., Abrikosov-fermion) representation of spin-$1/2$ operators.
In the experiment, the crystal structure up to 4 GPa belongs to the same space group as the material at ambient pressure and hence these spin liquid phases may be good candidates for non-trivial quantum paramagnetic ground states. The classification leads to 160 different $Z_2$ spin liquids, and 176 distinct $U(1)$ spin liquids. 
At the moment, there is no specific heat measurement under pressure and it is not clear whether there is a finite temperature transition or not in 2-4 GPa range.
If the ground state is a U(1) spin liquid, there will be no thermal transition, but only a crossover, while $Z_2$ spin liquid phases would show a finite temperature 
transition to a trivial paramagnetic state in three dimensions. 

Given that the Kitaev interaction is already significant at ambient pressure\cite{Biffin2014,Kim2016a,Katukuri2016}, we further investigate all possible ``root'' 
$U(1)$ spin liquid phases proximate to the Kitaev's $Z_2$ spin liquid state. 
This leads to 8 promising candidate $U(1)$ spin liquids for the high-pressure paramagnetic phase in $\beta$-Li$_2$IrO$_3$. 
If they occur in this material at high pressure, there will be no signature of finite temperature transition in future specific heat measurement.
We also discuss the physical properties of these 8 candidate states, and show that they all support one-dimensional nodal-line spectra on the spinon Fermi surface, similar to the hyperhoneycomb Kitave model\cite{Schaffer2015,OBrien2016}.

The rest of the paper is organized as follows. In section II, we provide the full classification of symmetric $Z_2$ and $U(1)$ spin liquid phases on the
hyperhoneycomb lattice.
We investigate the ``root'' $U(1)$ spin liquid states in proximity to the Kitaev $Z_2$ spin liquid in section III. The key properties of 8 ``root'' $U(1)$
spin liquid states are discussed in section IV. We conclude in section V.

	\section{Classification of U(1) and $ Z_2 $ Spin Liquids}
	
	In this section, we briefly introduce the framework of Projective Symmetry Group (PSG) classification for symmetric quantum spin liquids\cite{Wen2002}. We first review the lattice structure and space group symmetries of the hyperhoneycomb lattice, and then construct symmetric spin liquids in the Abrikosov-fermion representation.
	
	\subsection{Lattice structure and symmetries}
	\begin{figure}
		[h]
		\includegraphics[width=8cm]{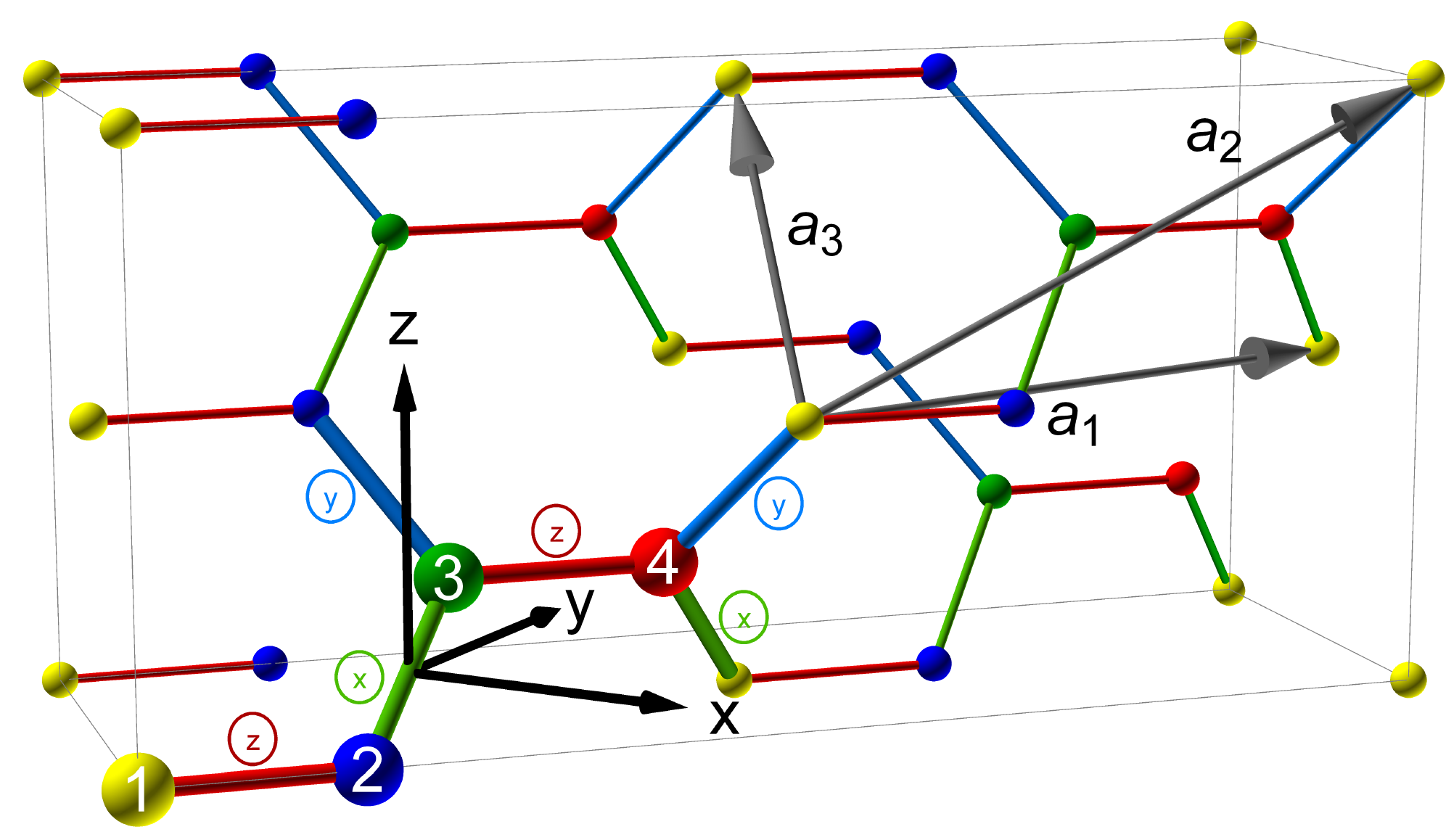}
		\caption{\label{fig:lattice} The hyperhoneycomb lattice.}
	\end{figure}
	
	The hyperhoneycomb lattice consists of 4 sublattices denoted by different colors in Fig. \ref{fig:lattice}. The Bravais lattice vectors are chosen as
	\begin{align}
	&\boldsymbol{a}_1 = (2,4,0), &\boldsymbol{a}&_2 = (3,3,2), &\boldsymbol{a}_3 = (-1,1,2).
	\label{avec}
	\end{align}
	The sublattice displacements are
	\begin{align}
	&\boldsymbol{d}_1 = \left(-1,-\frac{3}{2},-\frac{1}{2}\right), & \boldsymbol{d}&_2 = \left( 0, -\frac{1}{2}, -\frac{1}{2}\right), \nonumber \\
	&\boldsymbol{d}_3 = \left( 0, \frac{1}{2}, \frac{1}{2}\right), & \boldsymbol{d}&_4 = \left(1,\frac{3}{2},\frac{1}{2}\right),
	\end{align}
	where origin of the Cartesian coordinate system is chosen to be at the center of the nearest-neighbor (NN) bond connecting sublattices 2 and 3 within the same unit cell. A general lattice site can be labeled by
	\begin{equation}
	(n_1, n_2, n_3, s) = n_1\boldsymbol{a}_1 + n_2 \boldsymbol{a}_2 + n_3 \boldsymbol{a}_3 + \boldsymbol{d}_s.
	\label{sitelabel}
	\end{equation}
where $n_{1,2,3}\in\mathbb{Z}$ and $1\leq s\leq4$ is the sublattice index.
	
	There are 6 inequivalent NN bonds associated with one unit cell. They are divided into three classes, which are perpendicular to x- (green bonds), y- (blue bonds), and z- (red bonds) axes respectively:
	\begin{align}
	&\text{x-bonds:} &(&\mbox{sublattice }2\leftrightarrow 3): (0,1,1) \nonumber \\
	& &(&\text{sublattice }4\leftrightarrow 1+\boldsymbol{a}_1): (0,1,-1) \nonumber \\
	&\text{y-bonds:} &(&\mbox{sublattice }4\leftrightarrow 1+\boldsymbol{a}_2): (1,0,1) \nonumber \\
	& &(&\text{sublattice }3\leftrightarrow 2+\boldsymbol{a}_3): (-1,0,1) \nonumber \\
	&\text{z-bonds:} &(&\mbox{sublattice }1\leftrightarrow 2): (1,1,0) \nonumber \\
	& &(&\text{sublattice }3\leftrightarrow 4): (1,1,0)
	\label{xyzbonds}
	\end{align}

	In addition to Bravais lattice translations, the other generators of space group symmetries written in Cartesian coordinates are
	\begin{align}
	&\text{Inversion: } & (x,y,z)&\overset{\sigma}\longrightarrow (-x,-y,-z)\\
	&\text{Glide reflections:} & (x,y,z)&\overset{r_1}\longrightarrow (x,y,-z)+\boldsymbol{a}_1/2\\
	& & (x,y,z)&\overset{r_2}\longrightarrow (y,x,z)+\boldsymbol{a}_2/2,
	\end{align}
	plus the three translations along the Bravais lattice vectors. Alternatively, by labeling lattice sites in the format of (\ref{sitelabel}), the space group symmetries can be rewritten as
	\begin{enumerate}
		\item Translations
		\begin{align}
		&T_1: (n_1,n_2,n_3,s)\rightarrow (n_1+1,n_2,n_3,s)\\
		&T_2: (n_1,n_2,n_3,s)\rightarrow (n_1,n_2+1,n_3,s)\\
		&T_3: (n_1,n_2,n_3,s)\rightarrow (n_1,n_2,n_3+1,s)
		\end{align}
		\item Inversion $ \sigma $
		\begin{equation}
		\sigma: (n_1,n_2,n_3,s)\rightarrow (-n_1, -n_2, -n_3, \sigma(s)),
		\end{equation}
		where $\sigma(1)=4$, $\sigma(2)=3$, $\sigma(3)=2$, $\sigma(4)=1$.
		\item Glide reflection $r_1$
		\begin{multline}
		r_1: (n_1, n_2, n_3, s) \\
		\rightarrow (n_1+n_2+n_3+A_s,-n_3,-n_2,r(s)).
		\label{r1trans}
		\end{multline}
		\item Glide reflection $r_2$
		\begin{multline}
		r_2: (n_1, n_2, n_3, s) \\
		\rightarrow (-n_3, n_1+n_2+n_3+A_s,-n_1, r(s)).
		\label{r2trans}
		\end{multline}
		For both $ r_1 $ and $ r_2 $, $ A_s = 0 $ for $ s=1,2 $ and $ A_s=1 $ for $ s=3,4 $; the sublattice transforms are $ r(1)=3$, $r(2)=4$, $r(3)=1$, $r(4)=2 $.

	\end{enumerate}
	
	In summury, a general symmetry group symmetry operation can be expressed as
	\begin{equation}
	U={\cal T}^{\nu_{\cal T}} T_1^{\nu_{T_1}} T_{2}^{\nu_{T_2}} T_3^{\nu_{T_3}} \sigma^{\nu_\sigma}r_1^{\nu_{r_1}} r_2^{\nu_{r_2}},
	\end{equation}
	where $ \cal T $ is time reversal operation, $ \nu_{{\cal T},\sigma, r_1, r_2}\in \mathbb{Z}_2=\{0,1\}$, and $ \nu_{T_{1,2,3}}\in \mathbb{Z} $.
	
	\subsection{Projective symmetry groups}
	We briefly review the projective symmetry group (PSG) classification for symmetric spin liquid states\cite{Wen2002} in the Abrikosov-fermion representation\cite{Abrikosov1965a}. In order to construct a symmetric spin liquid ground state for an interacting spin model, we introduce the Abrikosov fermions (or ``slave fermions'')
	\begin{align}\label{spinfermion}
	&\Psi_{i} = \left(
	\begin{array}{cc}
	f_{i\uparrow} & f_{i\downarrow}^\dagger\\
	f_{i\downarrow} & -f_{i\uparrow}^\dagger
	\end{array}
	\right), & &\mathbf{S}_i = \frac{1}{4} \mbox{Tr}(\Psi_i^\dagger \boldsymbol{\sigma}\Psi_i),
	\\
	&\{f_{i\alpha}, f_{j\beta}^\dagger\}=\delta_{ij}\delta_{\alpha\beta},
	& &\{f_{i\alpha},f_{i\beta}\}=0=\{f_{i\alpha}^\dagger,f_{j\beta}^\dagger \}.
	\end{align}
	Here $ \mathbf{S}_i $ are spin-1/2 operators at site $ i $, and $ \boldsymbol{\sigma} $'s are Pauli matrices. The Abrikosov-fermion representation introduces a gauge redundancy for the ``slave fermions'' $\{f_{i\alpha}\}$ because annihilating a spin-up fermion $ f_{i\uparrow} $ has the same effect as creating a spin-down fermion $ f_{i\downarrow}^\dagger $ regarding the change of physical spin. Consequently a local gauge transformation can be performed by an SU(2) rotation $ W_i=e^{-i\boldsymbol{\phi}_i\cdot\boldsymbol{\sigma}/2} $ acting on the right as $ \Psi_i\rightarrow \Psi_iW_i $. One can directly see that such a transformation leaves the physical spin operator (\ref{spinfermion}) invariant. On the other hand, a physical spin rotation can be performed by acting an SU(2) rotation $ R=e^{-i\boldsymbol{\phi}_i\cdot\boldsymbol{\sigma}/2} $ on the left as $ \Psi_i\rightarrow R^\dagger_i\Psi_i $, which rotates $ \boldsymbol{S}_i $ by the Euler angle $ \boldsymbol{\phi}_i $.
	
	Within the Abrikosov-fermion representation, a bilinear spin-spin interaction in the Hamiltonian becomes 4-fermion interaction of (Arikosov) slave fermions, which upon the Hubbard-Stratonovich decomposition leads to a quadratic mean-field Hamiltonian of slave fermions. A generic mean field Hamiltonian can be expressed as
	\begin{equation}\label{Hmfgeneral}
	H_{MF} = \sum_{ij}\sum_{\mu=0,x,y,z} \mbox{Tr}(\sigma_\mu \Psi_i u_{ij}^{(\mu)}\Psi_j^\dagger),
	\end{equation}
	where $ \sigma_0 $ denotes a 2 by 2 identity matrix, and
	\begin{align}
	&u_{ij}^{(0)}=is_0\sigma_0 + \sum_{\mu=x,y,z} s_\mu \sigma_\mu ,\\
	&u_{ij}^{(x,y,z)}=t^{(x,y,z)}_0+ \sum_{\mu=x,y,z} i t^{(x,y,z)}_\mu \sigma_\mu
	\end{align}
	with $ \{ s_\mu, t^{(x,y,z)}_\mu|\, \mu=0,x,y,z\} $ being real numbers serving as mean field parameters. (Bond indices $ i,j $ are omitted here). One can verify that under a spin rotation, $ u_{ij}^{(0)} $ and $ (u_{ij}^{(x)},u_{ij}^{(y)}, u_{ij}^{(z)}) $ transform as scalars and vectors respectively, and represent the spin singlet and triplet terms for the mean field Hamiltonian. Within each $ u_{ij}^{(\mu)} $ matrix, components $ \sim \sigma_0, \sigma_3 $ represent hopping of fermionic spinons $ f_{i\alpha} $, while components $ \sim \sigma_1,\sigma_2 $ are associated with spinon pairings. The gauge redundancy is resolved after the Gutzwiller projection into the physical Hilbert space with one fermion per site, by enforcing the single-occupancy constraints $ f_{i\uparrow}^\dagger f_{i\uparrow}+f_{i\downarrow}^\dagger f_{i\downarrow}=1$, $f_{i\uparrow}^\dagger f_{i\downarrow}^\dagger=0 $. These constraints can be written in a compact form\cite{Wen2002}
	\begin{equation}
	\mathrm{Tr}(\Psi_i^\dagger \Psi_i\vec{\sigma})=0.
	\end{equation}
	
	To construct a symmetric spin liquid state without spontaneous symmetry breaking, the mean-field Hamiltonian should preserve all the symmetries of the spin models. However such a requirement can be loosen due to the gauge redundancy in the Abrikosov-fermion representation: the symmetry only needs to be preserved up to a gauge transformation. Specifically under the action of a symmetry operation $ U $, the fermionic spinons transform as
$U\Psi_iU^{-1} = R_U^\dagger \Psi_{U(i)}G_{U}(U(i))$, and hence the mean field Hamiltonian satisfies\cite{Huang2017}
	\begin{multline}
	\sum_{\mu} \mbox{Tr} \left[(R_{U}\sigma_\mu R_U^\dagger) \Psi_{i}G_U(i)\cdot u_{U^{-1}(i),U^{-1}(j)}^{(\mu)}\cdot G^\dagger_U(j)\Psi_{j}^\dagger \right] \\
	= \sum_{\mu } \mbox{Tr}\left(\sigma_\mu \Psi_{i} u_{ij}^{(\mu)} \Psi_j^\dagger \right).
	\label{psggeneral}
	\end{multline}
	Here $R_U$ and $\{G_U(i)\}$ are $SU(2)$ matrices for spin rotations and local gauge transformations respectively. In particular, the subgroup $ G_{\boldsymbol{e}} $ for gauge transformations associated with the identity element $ \boldsymbol{e} $ of the symmetry group is called the invariant gauge group (IGG). The PSG can be regarded as an extension of the spatial symmetry group (SG):  SG=PSG/IGG. The different choice of IGG corresponds to spin liquids with different quantum orders\cite{Wen2002}: e.g. IGG$=\{\pm1\}$ in a $Z_2$ spin liquid, IGG$=\{e^{-i\phi\sigma_z/2}|0\leq\phi<4\pi\}\simeq U(1)$ in a $U(1)$ spin liquid, and IGG$=\{e^{-i\boldsymbol{\phi}\cdot\boldsymbol{\sigma}/2}\}\simeq SU(2)$ for an $SU(2)$ spin liquid. The choice of gauge group elements $ G_U $ is not unique because of the following gauge rotations:
	\begin{align}
	G_U(i) &\rightarrow W_i G_U(i) W_{U^{-1}(i)}^\dagger, \qquad W_i \in SU(2), \\ \label{mftransform}
	u_{ij} &\rightarrow W_i u_{ij} W_j^\dagger, \qquad \Psi_i \rightarrow\Psi_i W^\dagger_i.
	\end{align}
	leaves the mean-field ansatz and above symmetry condition (\ref{psggeneral}) invariant. The gauge-inequivalent choices of gauge transformations $\{G_U(i)\}$ can be determined by requiring its compatibility with the group structure of SG. Solving the algebraic equations from the symmetry group structure can lead to different PSGs, or different extensions of the same symmetry group SG. They correspond to distinct quantum spin liquid states with different physical properties.

It turns out there exist {\bf 160 $ Z_2 $} spin liquids, and {\bf 176 U(1)} spin liquids on the hyperhoneycomb lattice. Details of the PSG classification can be found in Appendix \ref{app:psg}. The results of $ Z_2 $ PSGs are detailed in Appendix \ref{app:z2result}. For U(1) PSGs, the gauge transformations associated with spatial symmetries are given in Appendix \ref{app:u1result_spatial}, while the gauge transformation associated with time reversal symmetry is given by Eqs.~(\ref{app:u1result_t1}), (\ref{app:u1result_t2_eq1})--(\ref{app:u1result_t2_eq8}).

\section{Root U(1) spin liquids in proximity to the Kitaev spin liquid on hyperhoneycomb lattice}

The solvable Kitaev model is argued to be in close proximity to a class of magnetic ``Kitaev materials'' with strong spin-orbit couplings, including the hyperhoneycomb iridate $\beta$-Li$_2$IrO$_3$\cite{Biffin2014,Takayama2015}. This makes ``Kitaev spin liquid'', the ground state of Kitaev model, a reasonable starting point to understand possible spin liquid phases realized in hyperhoneycomb iridates.

In this section, we first work out the PSGs of Kitaev spin liquid, which encodes how time reversal and spatial symmetries are implemented on the fractionalized exctitations therein. Next, we identify 8 ``root'' $U(1)$ spin liquid states, all of which neighbor the Kitaev spin liquid by a continuous Higgs transition, which breaks the IGG from $U(1)$ down to $Z_2$. These 8 $U(1)$ spin liquids are promising candidates for the possible spin liquid phase observed in pressurized $\beta$-Li$_2$IrO$_3$.

	\subsection{PSG for the Kitaev spin liquid on hyperhoneycomb lattice}
\begin{figure}
		[h]
		\centering
		\includegraphics[width=4cm]{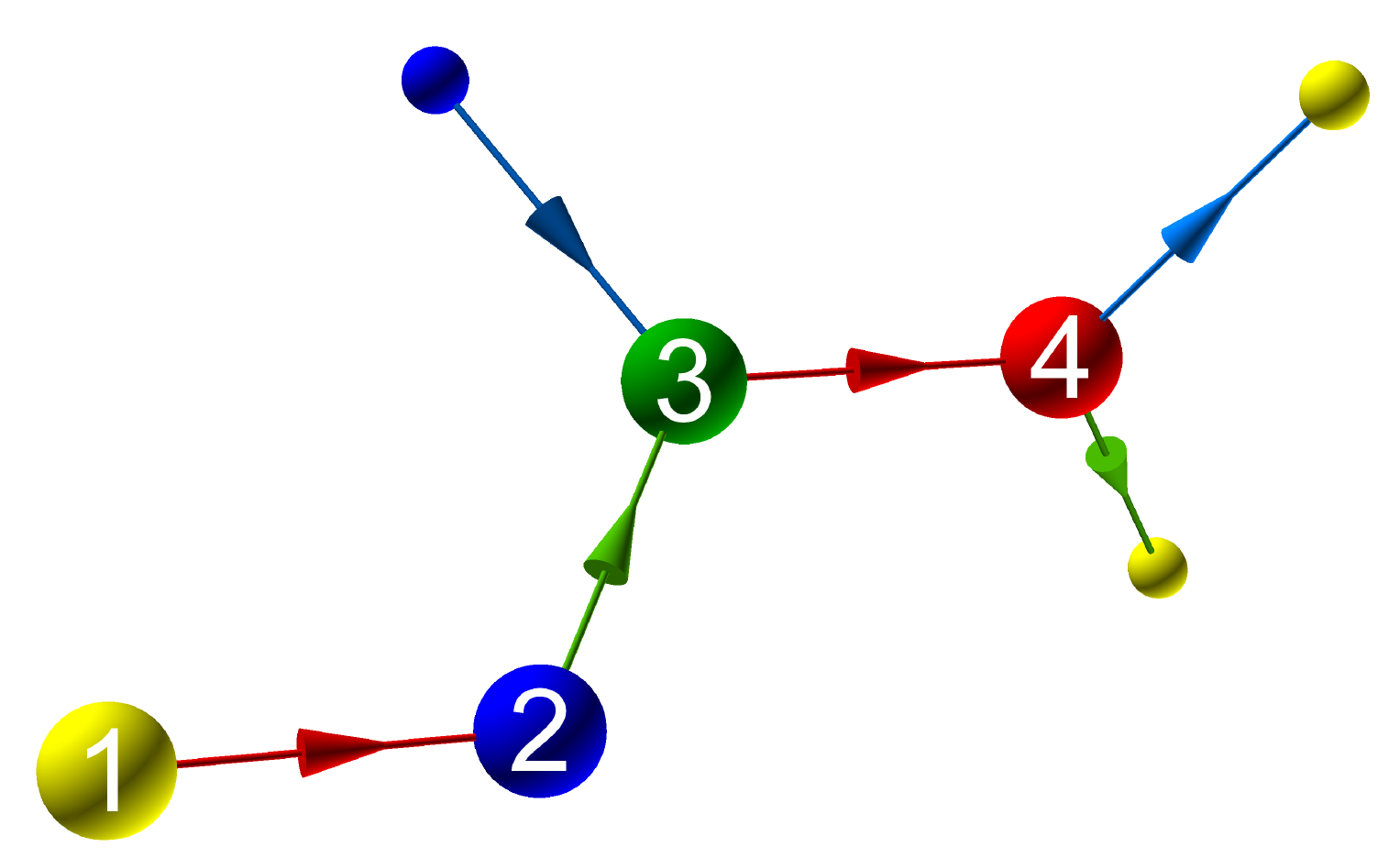}\qquad
		\includegraphics[width=4cm]{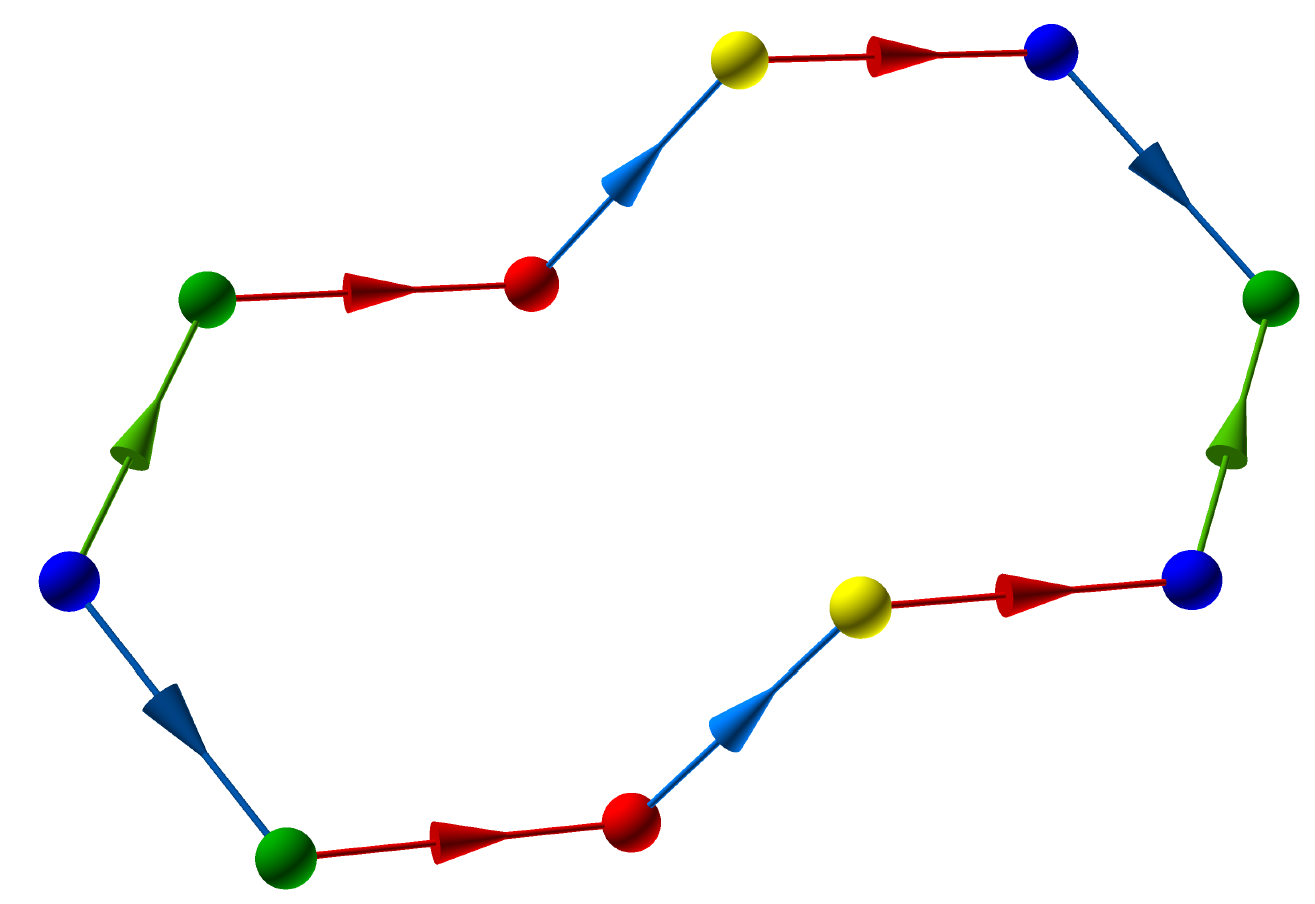}
		\caption{\label{fig:kitaev} The directions of $ +iK_{ij}^{(\alpha)} $ in the ground state of Kitaev model with all $ J^{(\alpha)} > 0 $. It has zero flux around any closed loop.}
	\end{figure}

	The Kitaev model on any trivalent lattice
	\begin{equation}
	H_K=\sum_{\langle ij\rangle \in\alpha} J^{(\alpha)} S_i^\alpha S_j^\alpha
	\end{equation}
	involves bond-dependent Ising-type interactions between nearest neighbors. On the hyperhoneycomb lattice, $ \alpha=x,y,z $ for the x-, y-, and z-bonds specified in Eq.~(\ref{xyzbonds}). The above Kitaev model can be exactly solved\cite{Kitaev2006} in terms of Majorana fermions $ (c_i, b_i^x, b_i^y, b_i^z)$,
	\begin{align}
	&S_i^\alpha = ib_i^\alpha c_i, \quad \{c_i,c_j\}=2\delta_{ij},~ \{b_i^\alpha, b_j^\beta\}=2\delta_{ij}\delta_{\alpha\beta}, \label{kitaevcc}\\
	&
	H_K = \sum_{\langle ij\rangle\in\alpha} iK^{(\alpha)}_{ij} c_i c_j, \quad K_{ij}^{(\alpha)} = -iJ^{(\alpha)}b_i^\alpha b_j^\alpha.\label{kitaev_ansatz}
	\end{align}
	The link variables $\{K_{ij}^{(\alpha)}\}$ commute with the Hamiltonian and with each other, and therefore are integrals of motion. Note $ (K_{ij}^{(\alpha)})^2=(J^{(\alpha)})^2 $, and therefore $ K_{ij}^{(\alpha)}=-K_{ji}^{(\alpha)}= \pm J^{(\alpha)}$. In the ground state, in order to minimize the energy of the above $\{c_i\}$-fermion hopping model, the link variables $ iK_{ij}^{(\alpha)} $ on a hyperhoneycomb lattice are shown to support zero flux in any closed loop\cite{Schaffer2015}. We therefore choose a sign convention of $\{K_{ij}^{(\alpha)}\}$ such that they are positive along sublattices $ (1\rightarrow2\rightarrow3\rightarrow4\rightarrow1) $, as shown in Fig.~\ref{fig:kitaev}. Such a configuration is invariant under the primitive lattice translations.
	
	Our goal is to find the projective symmetry group of the Kitaev spin liquid ans\"{a}tze in Eq.~(\ref{kitaevcc}), with a given zero-flux configuration $ iK_{ij}^{(\alpha)}$. First, we notice that the Majorana fermions can also be written in terms of complex slave fermions,
	\begin{align}
	& f_{i\uparrow} = b_i^z + ic_i, & f_{i\downarrow} = b_i^x + ib_i^y.
	\end{align}
	Compared with (\ref{spinfermion}), one can easily see that
	\begin{equation}
	\Psi_i = ic_i\sigma_0 + \sum_{\mu=x,y,z} b_i^\mu \sigma_\mu
	\end{equation}
	Hence under a symmetry operation $U$, the Majorana fermions transform as
	\begin{multline}
	\Psi_i \rightarrow R_U^\dagger \Psi_{U(i)} G_U(U(i)) = ic_{U(i)} \left( R_U^\dagger G_U(U(i))\right) \\
	+ \sum_{\alpha=x,y,z} b_{U(i)}^\alpha \left( R_U^\dagger \sigma_\alpha G_U(U(i)) \right).\label{majorana sym}
	\end{multline}
	
	Here $ R_U,\{G_U(i)\}$ are spin and associated gauge SU(2) rotations respectively. Clearly the above Kitaev spin liquid has a $ Z_2 $ IGG generated by the following global gauge rotation:
\begin{equation}
c_i\rightarrow-c_i,~~b_i^\mu\rightarrow-b_i^\mu,~~~\forall~i,\mu.
\end{equation}
which leaves the Majorana hopping ansatz invariant. In particular the Kitaev spin liquid ansatz (\ref{kitaev_ansatz}) has one special feature: while all $\{b^\mu_i\}$ fermions are dimerized on one NN bond, the $\{c_i\}$ fermions are delocalized and can hop around the entire lattice. The sharp difference between $\{b^\mu_i\}$ and $\{c_i\}$ fermions indicate that they cannot be mixed by any symmetry transformation in the Kitaev spin liquid. This unusual property, together with symmetry implementation (\ref{majorana sym}) dictates that for any symmetry operation $U$ preserved by the Kitaev spin liquid, we must have
%that the signs of $ K_{ij}^{(\alpha)} $ in Eq.~(\ref{kitaevcc}) can be flipped for a given type of bond $ \alpha $ within a closed loop. As one can verify that a closed loop in hyperhoneycomb lattices always involve an even number of bonds per type $ \alpha $, such a flip leaves the flux through a loop unchanged. That means after a gauge transformation, the resulting quadratic Hamiltonian Eq.~(\ref{Hmfgeneral}) should still be expressed in terms of $ c_i $'s. Thus, we need
	\begin{equation}\label{majorana sign}
	R_U^\dagger G_U(i) = \eta_U^i\tau_0 \Leftrightarrow G_U(i) = \eta_U^i R_U,~~~\eta_U=\pm1.
	\end{equation}
in order for the Kitaev ansatz (\ref{kitaev_ansatz}) to be invariant under symmetry $U$. In particular $\eta_U=-1$ for every symmetry $U$ that reverses the NN Majorana hopping sign in Fig. \ref{fig:kitaev}, where the sublattice-dependent sign structure in (\ref{majorana sign}) brings back the original Kitaev ansatz.

As a result, the gauge rotation $\{G_U(i)\}$ associated with symmetry operation $U$ is fixed by its corresponding spin rotation $R_U$ up to a sublattice sign (\ref{majorana sign}). For the Kitaev spin liquid state on the hyperhoneycomb lattice, the SU(2) spin rotations associated with spatial symmetry operations are given by (denote $ \sigma_0\equiv\mathbb{I}_{2\times2} $, the identity matrix, and $ \sigma_{1,2,3}\equiv\sigma_{x,y,z} $)
	\begin{align}\nonumber
	&R_{T_1,T_2,T_3}^\dagger = \sigma_0, & R&_\sigma^\dagger =\sigma_0, \\
	&R_{r_1}^\dagger = i\sigma_3, & R&_{r_2}^\dagger= \frac{i}{\sqrt2}(\sigma_1-\sigma_2).
	\end{align}
	%Now, note the PSG condition Eq.~(\ref{psggeneral}), and the sign convention for the mean field ansats, we have
And their associated gauge transformations are given by
	\begin{align}\nonumber
	&G_{1,2,3}=\sigma_0, &G&_{r_1}=i\sigma_3,  &G&_{r_2}=\frac{i}{\sqrt2} (\sigma_1-\sigma_2),\\
	&G_\sigma(1,3)=\sigma_0, &G&_\sigma(2,4)=-\sigma_0. & &
	\label{kitaevpsg1}
	\end{align}
since $\eta_{T_{1,2,3}}=\eta_{r_{1,2}}=+1$ and $\eta_\sigma=-1$.
	Finally, time reversal symmetry must satisfy the following condition
	\begin{equation}\label{trequirement}
	G_{\cal T}(i)u_{ij}^{(\mu)}G_{\cal T}^\dagger (j) = -u_{ij}^{(\mu)},
	\end{equation}
	and in Kitaev spin liquid with only NN hoppings, we can choose
	\begin{align}\label{kitaevpsg2}
	&G_{\cal T}(1,3)=\sigma_0, &G_{\cal T}(2,4)=-\sigma_0.
	\end{align}
	Eqs. (\ref{kitaevpsg1}) and (\ref{kitaevpsg2}) summarize the PSG for Kitaev spin liquid (\ref{kitaev_ansatz}) on the hyperhoneycomb lattice.

	\subsection{Root U(1) states of Kitaev spin liquid}

Next we identify the ``root'' $U(1)$ spin liquid states in proximity to solvable Kitaev spin liquid. More concretely, we look for $U(1)$ PSGs whose symmetry implementations are compatible with the Kitaev spin liquid ansatz (\ref{kitaev_ansatz}). These $U(1)$ states can be tuned into the Kitaev spin liquid through a Higgs transition, during which the spinon pairing terms break the gauge group from $U(1)$ down to $Z_2$. Here we will skip the details of the calculation, which are given in Appendix \ref{app:kitaevPSG}. There turns out to be 8 $U(1)$ root states for the Kitaev spin liquids. Constrained by the nonsymmorphic space group, all of them share the trivial gauge transformation associated with lattice translation symmetries $T_{1,2,3}$:
	\begin{equation}
	G_{1,2,3}=\sigma_0.
	\end{equation}
	For all 8 root $U(1)$ states, the gauge transformation associated with time reversal symmetry is also the same as in Kitaev spin liquids
	\begin{align}\label{U1kitaevT}
	&G_{\cal T}(1,3)=\sigma_0, &G_{\cal T}(2,4)=-\sigma_0.
	\end{align}
	For the remaining space group symmetries, their associated gauge transformations are summarized in Table \ref{tab:U1kitaev}. In all cases, the gauge rotations only depend on sublattice index as $ G_U(s=1,2,3,4) $, but are independent of the unit cell index $(x,y,z)$.
	\begin{table}
		[h]
		\caption{\label{tab:U1kitaev} The PSGs of eight $U(1)$ root states for the Kitaev spin liquid on a hyperhoneycomb lattice. The gauge transformations $ G_{T_{1,2,3}} $ for lattice translations are all trivially $ \sigma_0 $, and $G_{\cal T} $ for time reversal is given by Eq.~(\ref{U1kitaevT}).}
		\begin{tabular}{|c|c|c|c|c|c|c|c|}
			\hline
			\# & $G_\sigma(1,4)$ & $G_\sigma(2,3)$ & $G_{r_1}(s)$ & $G_{r_2}(1)$ & $G_{r_2}(2)$ & $G_{r_2}(3)$ & $G_{r_2}(4)$ \\
			\hline
			1 & $i\sigma_3$ & $i\sigma_3$ & $i\sigma_3$ & $\sigma_0$ & $-\sigma_0$ & $-\sigma_0$ & $\sigma_0$ \\
			\hline
			2 & $ i\sigma_3 $ & $ -i\sigma_3 $ & $ i\sigma_1 $ & $ i\sigma_3 $ & $ -i\sigma_3 $ & $ i\sigma_3 $ & $ -i\sigma_3 $ \\
			\hline
			3 & $ i\sigma_3 $ & $ i\sigma_3 $ & $ i\sigma_3 $ & $ i\sigma_1 $ & $ -i\sigma_1 $ & $ i\sigma_1 $ & $ -i\sigma_1 $ \\
			\hline
			4 & $ i\sigma_1 $ & $ -i\sigma_1 $ & $ \sigma_0 $ & $ i\sigma_3 $ & $ i\sigma_3 $ & $ -i\sigma_3 $ & $ -i\sigma_3 $ \\
			\hline
			5 & $ i\sigma_1 $ & $ i\sigma_1 $ & $ i\sigma_1 $ & $ i\sigma_3 $ & $ -i\sigma_3 $ & $ i\sigma_3 $ & $ -i\sigma_3 $ \\
			\hline
			6 & $ i\sigma_1 $ & $ -i\sigma_1 $ & $ i\sigma_3 $ & $ i\sigma_1 $ & $ i\sigma_1 $ & $ -i\sigma_1 $ & $ -i\sigma_1 $ \\
			\hline
			7 & $ i\sigma_3 $ & $ -i\sigma_3 $ & $ i\sigma_2 $ & $ i\sigma_1 $ & $ i\sigma_1 $ & $ i\sigma_1 $ & $ i\sigma_1 $ \\
			\hline
			8 & $ i\sigma_1 $ & $ i\sigma_1 $ & $ i\sigma_1 $ & $ i\sigma_2 $ & $ -i\sigma_2 $ & $ i\sigma_2 $ & $ -i\sigma_2 $ \\
			\hline
		\end{tabular}
	\end{table}

\section{Properties of $U(1)$ spin liquids on hyperhoneycomb lattice}

Among the 160 symmetric $Z_2$ spin liquids and 176 symmetric $U(1)$ spin liquids on the hyperhoneycomb lattice, here we focus on the 8 $U(1)$ root states in proximity to the $Z_2$ spin liquid ground state of the solvable Kitaev model. We discuss the mean-field ansatz and physical properties of these 8 states, and in particular identify topologically protected nodal rings in the spinon spectrum of all 8 states. Experimentally, these 8 states are promising candidates for the possible spin liquid phase realized in $ \beta $-Li$ _2 $IrO$ _3 $ under high pressure.

%	The high symmetry of the hyperhoneycomb lattice results in the large number of symmetric spin liquid states. In the remaining part of this paper, we focus on a special class of U(1) states that serves as root states for Kitaev spin liquids in hyperhoneycomb lattices. These U(1) states hold PSG's compatible with that for the $ Z_2 $ Kitaev spin liquids, and are smoothly connected to the Kitaev spin liquids through the Anderson Higgs mechanism. Experimentally, these states are candidate U(1) spin liquids when the spin interaction is driven away from the limit where Kitaev interaction dominates, such as $ \beta $-Li$ _2 $IrO$ _3 $ under high pressure.

	\subsection{Mean-field ansatz of 8 root $U(1)$ spin liquids}
\begin{figure}
		[h]
		\includegraphics[width=5cm]{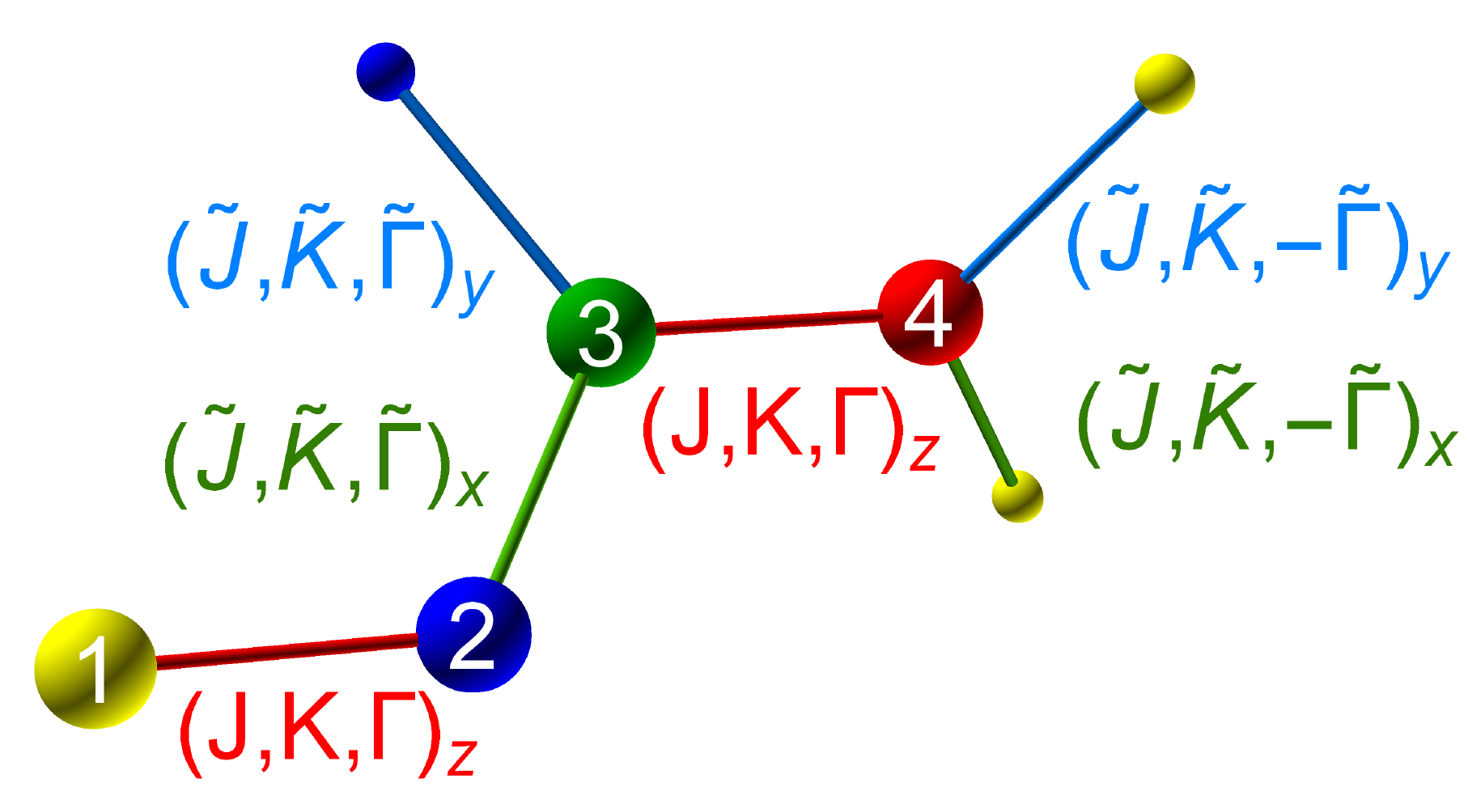}
		\caption{\label{fig:spin} The nonzero parameters and their relations in different bonds for the spin model Eq.~(\ref{spinmodel}). }
	\end{figure}

	The physical spin model describing the magnetism of $ \beta\mbox{-Li}_2\mbox{IrO}_3 $ was argued to take the following form~\cite{Rau2014,Winter2017}
	\begin{equation}\label{spinmodel}
	H=\sum_{ij\in\gamma}
	J_{ij}\mathbf{S}_i\cdot \mathbf{S}_j +K_{ij}^{(\gamma)} S_i^\gamma S_j^\gamma + \Gamma_{ij}^{(\gamma)} (S_i^\alpha S_j^\beta + S_i^\beta S_i^\alpha).
	\end{equation}
	Here $ (\alpha,\beta,\gamma)\sim (x,y,z) $, and $ J_{ij}, K_{ij}, \Gamma_{ij} $ are the Heisenberg, Kitaev, and symmetric anisotropy interactions respectively. Note that the symmetry of hyperhoneycomb lattice has already excluded interactions of Dzyaloshinskii-Moriya type $ \mathbf{D}_{ij}\cdot (\mathbf{S}_i\times \mathbf{S}_j) $. Further material considerations together with space group symmetries indicate that there are only 6 free parameters: $ (J, K, \Gamma) $ on z-bonds and $ (\tilde{J}, \tilde{K}, \tilde{\Gamma}) $ on x- and y-bonds, as illustrated in Fig.~\ref{fig:spin}. Here the subscript $ \gamma=x,y,z $ means that there is only one nonvanishing $ K_{ij}^{(\gamma)}$ and $\Gamma_{ij}^{(\gamma)} $ on each $\gamma-$type link $\langle ij\rangle$.  For instance, in bond $ (1,2) $ we have $ K_{12}^{(z)}=K,~~\Gamma_{12}^{(z)}=\Gamma $, while $ K_{12}^{(x,y)}=\Gamma_{12}^{(x,y)}=0 $.
	
	Starting from the spin model (\ref{spinmodel}), within the Abrikosov-fermion representation (\ref{spinfermion}) one can perform the following mean-field decomposition
	\begin{align}
	&\rho_s = \frac{\langle f_{i\uparrow}^\dagger f_{j\uparrow} \rangle + \langle f_{i\downarrow}^\dagger f_{j\downarrow}\rangle}{2},
	&\rho&_z = \frac{\langle f_{i\uparrow}^\dagger f_{j\uparrow} \rangle - \langle f_{i\downarrow}^\dagger f_{j\downarrow}\rangle}{2i},\\
	&\rho_x = \frac{\langle f_{i\uparrow}^\dagger f_{j\downarrow}\rangle - \langle f_{i\downarrow}^\dagger f_{j\uparrow}\rangle}{2i},
	&\rho&_y = \frac{\langle f_{i\uparrow}^\dagger f_{j\downarrow}\rangle + \langle f_{i\downarrow}^\dagger f_{j\uparrow}\rangle}{2},
	\end{align}
	and obtain the mean-field ans\"{a}tze (\ref{Hmfgeneral}). Note in the canonical gauge for U(1) spin liquids~\cite{Wen2002}, only hopping terms are present. Denote the mean field amplitudes in bond $ (1,2) $ as $ (s_\mu, t^{(x,y,z)}_\mu) $, in bond $ (2,3) $ as $ (\tilde{s}_\mu, \tilde{t}^{(x,y,z)}_\mu) $, the generic spin Hamiltonian (\ref{spinmodel}) yields $ s_0=t^x_0 =t^y_0=t^z_0 = 0 $ and $ \tilde s_0=\tilde t^x_0 =\tilde t^y_0=\tilde t^z_0 = 0 $ in the mean-field decoupling. The remaining parameters $ (s_3, t^{(x,y,z)}_3) $ and $ (\tilde{s}_3, \tilde{t}^{(x,y,z)}_3) $ are further constrained by requirement (\ref{psggeneral}) for any symmetric spin liquids. Here we focus on the root U(1) states of the Kitaev spin liquid specified by PSGs in Table \ref{tab:U1kitaev}, and write down the mean-field ansatz of them all.

	To analyze the spinon band structures of U(1) spin liquids, it is convenient to adopt the following basis in momentum space
	\begin{equation}
	\Phi_{\boldsymbol{k}} = (f_{\boldsymbol{k}1\uparrow}, f_{\boldsymbol{k}1\downarrow}, f_{\boldsymbol{k}3\uparrow}, f_{\boldsymbol{k}3\downarrow}, f_{\boldsymbol{k}2\uparrow}, f_{\boldsymbol{k}2\downarrow}, f_{\boldsymbol{k}4\uparrow}, f_{\boldsymbol{k}4\downarrow})^T
	\end{equation}
	where $ T $ means transpose. The time reversal symmetry (\ref{U1kitaevT}) forbids onsite terms in the mean-field ansatz. Moreover the bipartite NN couplings only hop spinons between odd sublattices $s=1,3$ and even sublattices $s=2,4$. Therefore the mean-field Hamiltonian takes the off-diagonal form
	\begin{align}\label{hmfD}
	&H_{MF} = \sum_{\boldsymbol{k}} \Phi_{\boldsymbol{k}}^\dagger {\cal H}_{\boldsymbol{k}} \Phi_{\boldsymbol{k}}, &
	{\mathcal H}_{\boldsymbol{k}} = \left(
	\begin{array}{cc}
	0 & D_{\boldsymbol{k}}\\
	D_{\boldsymbol{k}}^\dagger & 0
	\end{array}
	\right).
	\end{align}
featuring the sublattice symmetry
\begin{align}\label{chiral sym}
&\{{\mathcal H}_{\boldsymbol{k}},\tau^z\}=0, &\tau^z\equiv\left(
	\begin{array}{cc}
	\mathbb{I}_{4\times4} & 0\\
	0 &-\mathbb{I}_{4\times4}.
	\end{array}
	\right)
\end{align}
	Each 4 by 4 matrix $ D_{\boldsymbol{k}} $ can be written as
	\begin{equation}\label{dmat}
	D_{\boldsymbol{k}} = \left(
	\begin{array}{cc}	
	h_{12}^\dagger & h_{41_x}e^{ik_1} + h_{41_y} e^{ik_2}\\
	h_{23} + h_{2_z3}e^{-ik_3} & h_{34}^\dagger
	\end{array}
	\right),
	\end{equation}
	where $ k_\mu = \boldsymbol{k}\cdot \boldsymbol{a}_\mu $ with $ \boldsymbol{a}_\mu $ the three Bravais lattice vectors (\ref{avec}). Here we denote the nonzero free parameters $ (s_3, t^{(x,y,z)}_3)\equiv (s, t^{(x,y,z)}) $ in bond $ (1,2) $, and similarly $ (\tilde{s}_3, \tilde{t}^{(x,y,z)}_3)\equiv (\tilde{s}, \tilde{t}^{(x,y,z)}) $ in bond $ (2,3) $,
	\begin{align}\nonumber
	&h_{12}=h(s, t^x, t^y, t^z), \quad h_{23}=h(\tilde{s}, \tilde{t}^x, \tilde{t}^y, \tilde{t}^z),\\
	&h(s, t^x, t^y, t^z) \equiv \left(
	\begin{array}{cc}
	-s+it^z & t^y+it^x\\ \label{hmatrix}
	-t^y + it^x & -s -it^z
	\end{array}
	\right).
	\end{align}
	These parameters together with $ h_{\mu\nu} $'s are subject to further constraints by the projective symmetry condition (\ref{psggeneral}), for different PSGs in Table \ref{tab:U1kitaev}. We summarize the NN mean field Hamiltonians for the 8 root $U(1)$ states in Table \ref{tab:U1kitaev}:
	\begin{enumerate}
		\item Parameters $ \{ \tilde{s}, t^x \} $:
		\begin{align}\nonumber
		&h_{12}=h(0,t^x,-t^x,0), &h&_{23}=h(\tilde{s},0,0,0),\\  \label{hcase1}
		&h_{34}=-h_{12}, &h&_{2_z3}=h_{41_x}=h_{41_y}=h_{23}.
		\end{align}
		
		\item Parameters $ \{ s, \tilde{s}, t^x  \} $:
		\begin{align}
		\nonumber
		&h_{12} = h(s,t^x, t^x, 0), &h&_{23}=h(\tilde{s}, 0,0,0),\\ \label{hmf2}
		&h_{34}= h(-s, t^x, t^x,0), &h&_{2_z3}=-h_{41_x}=-h_{41_y}=h_{23}.
		\end{align}

		\item Parameters $ \{ s, \tilde{s}, t^x \} $:
		\begin{align}
		\nonumber
		&h_{12}=h(s, t^x, t^x, 0), &h&_{23}=h(\tilde{s}, 0,0,0), \\
		&h_{34}=h(s,-t^x, -t^x,0), &h&_{2_z3}=h_{41_x}=h_{41_y}=h_{23}.
		\end{align}

		\item Parameters $ \{ s,t^x, \tilde{t}^x, \tilde{t}^y, \tilde{t}^z \} $:
		\begin{align}\nonumber
		&h_{12}=h(s,t^x,t^x,0), &h&_{23}=h(0,\tilde{t}^x, \tilde{t}^y, \tilde{t}^z),\\ \nonumber
		&h_{34}=h(s,-t^x,-t^x,0), &h&_{2_z3}= h(0,-\tilde t^y,  -\tilde t^x, \tilde t^z), \\
		&h_{41_x}= h(0,-\tilde{t}^x, -\tilde{t}^y, \tilde{t}^z), &h&_{41_y}=h(0, \tilde{t}^y, \tilde{t}^x, \tilde{t}^z).
		\end{align}
		
		\item Parameters $ \{ s, t^x, \tilde{t}^x, \tilde{t}^y, \tilde{t}^z \} $:
		\begin{align}
		\nonumber
		&h_{12}=h(s,t^x,t^x,0), &h&_{23}= h(0,\tilde{t}^x,\tilde{t}^y,\tilde{t}^z),
		\\ \nonumber
		&h_{34}=h(-s, t^x, t^x,0), &h&_{2_z3}=h(0,\tilde{t}^y, \tilde{t}^x, -\tilde{t}^z),\\
		&h_{41_x}=h(0, \tilde{t}^x, \tilde{t}^y, \tilde{t}^z), &h&_{41_y}=h(0, \tilde{t}^y, \tilde{t}^x, \tilde{t}^z).
		\end{align}		
		
		\item Parameters $ \{ s,t^x, \tilde{t}^x, \tilde{t}^y, \tilde{t}^z  \} $:
		\begin{align}
		\nonumber
		&h_{12}=h(s, t^x, t^x, 0), &h&_{23}=h(0, \tilde{t}^x, \tilde{t}^y, \tilde{t}^z),\\
		\nonumber
		&h_{34}=h(s, -t^x, -t^x,0), &h&_{2_z3} = h(0, \tilde{t}^y, \tilde{t}^x, -\tilde{t}^z),\\
		&h_{41_x}=h(0,-\tilde{t}^x, -\tilde{t}^y, \tilde{t}^z), &h&_{41_y}=h(0, -\tilde{t}^y, -\tilde{t}^x, -\tilde{t}^z).
		\end{align}
		
		\item Parameters $ \{ s, \tilde{s}, \tilde{t}^x  \} $:
		\begin{align}
		\nonumber
		&h_{12}=h(s, t^x, t^x, 0), &h&_{23}=h(\tilde{s}, 0,0,0),\\
		&h_{34}=h(-s, t^x, t^x, 0), &h&_{2_z3}=-h_{41_x}=-h_{41_y}=h_{23}.
		\end{align}
		Due to the specific model (\ref{spinmodel}), up to nearest neighbor terms this Hamiltonian is the same as case \#2, Eq.~(\ref{hmf2}).
		
		\item Parameters $ \{ t^x, \tilde{t}^x, \tilde{t}^y, \tilde{t}^z  \} $:
		\begin{align}\nonumber
		&h_{12}=h(0,t^x,-t^x,0), &h&_{23}=h(0,\tilde{t}^x, \tilde{t}^y, \tilde{t}^z),\\ \nonumber
		&h_{34}=h_{12}, &h&_{2_z3}=h(0, -\tilde{t}^y, -\tilde{t}^x, \tilde{t}^z),\\ \label{hcase8}
		&h_{41_x}=h(0, \tilde{t}^x,\tilde{t}^y, -\tilde{t}^z), &h&_{41_y}=h(0, -\tilde{t}^y, -\tilde{t}^x, -\tilde{t}^z).
		\end{align}
		
	\end{enumerate}
	
	\subsection{Topological spinon nodal rings}
\begin{figure*}
		[t]
		\parbox{4.7cm}{(a1)}
		\parbox{4.8cm}{(b1)}
		\parbox{7cm}{(c1)}\\
		\parbox{4.7cm}{ \quad \\
			\centering
		\includegraphics[width=4.0cm]{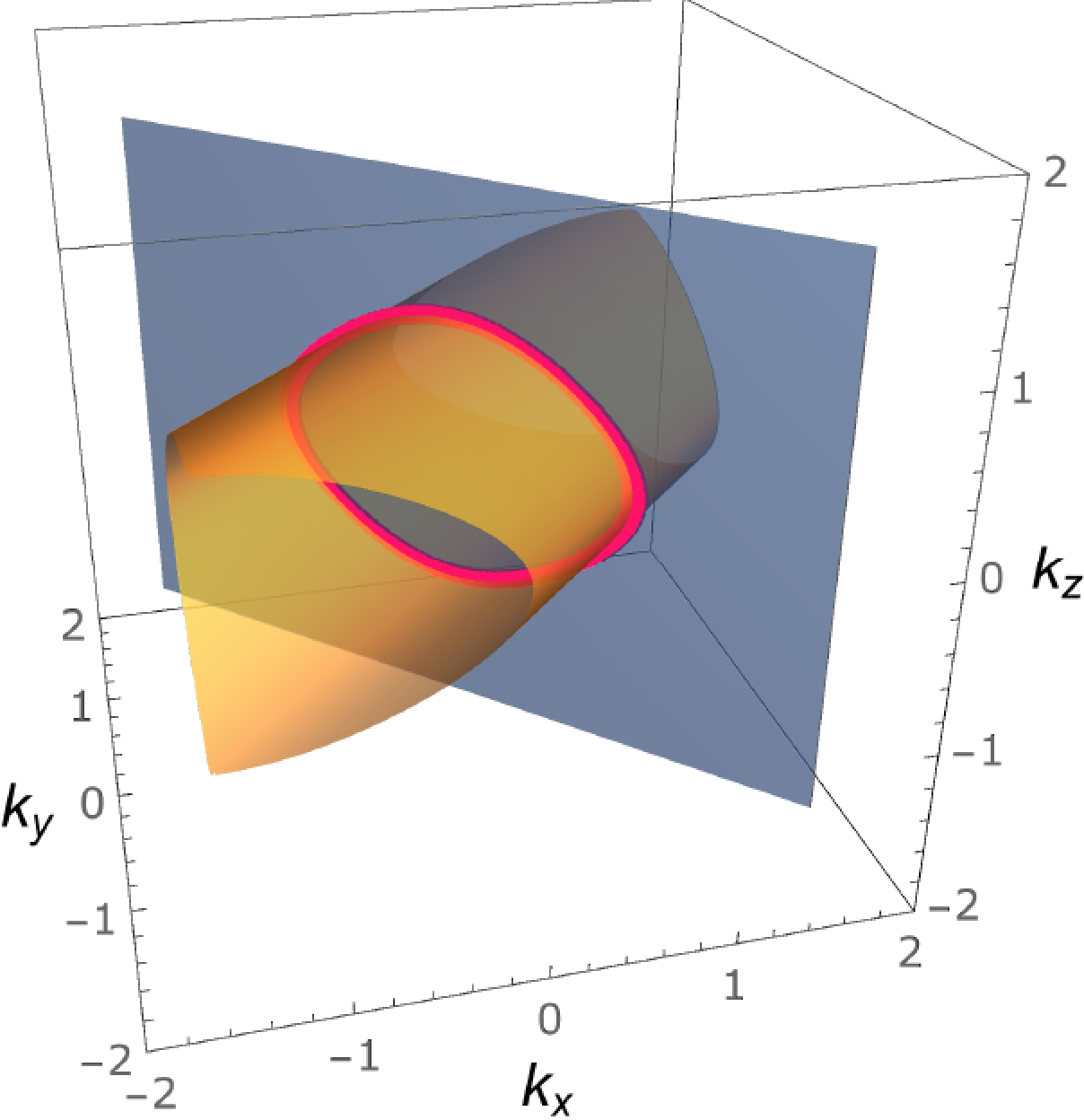}
	}
		\parbox{4.8cm}{ \quad \\\centering
		\includegraphics[width=4.2cm]{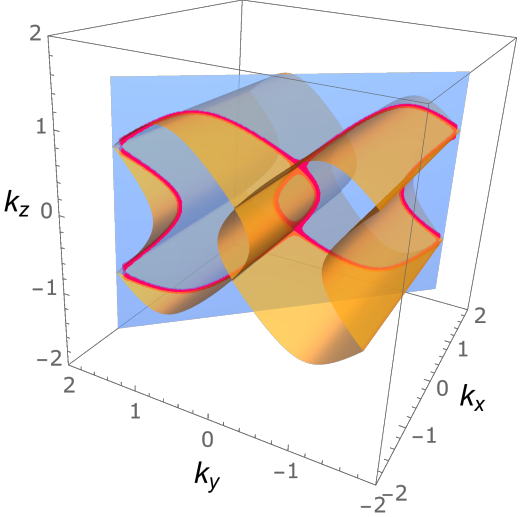}
	}
		\parbox{7cm}{ \centering
			\begin{picture}(50,135)
			\put(-60,10){\includegraphics[width=3.8cm]{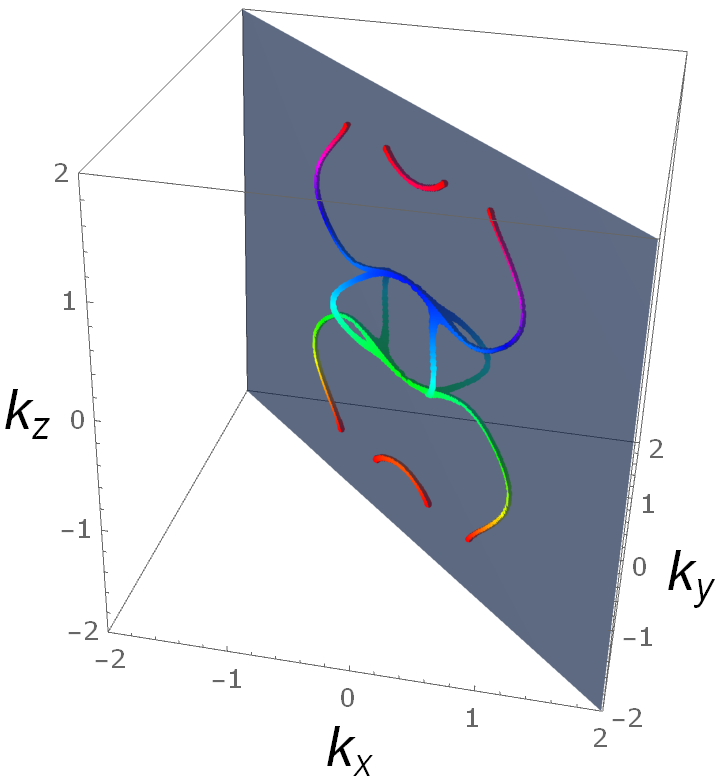} }
			\put(45,45){\boxed{\includegraphics[width=2.5cm]{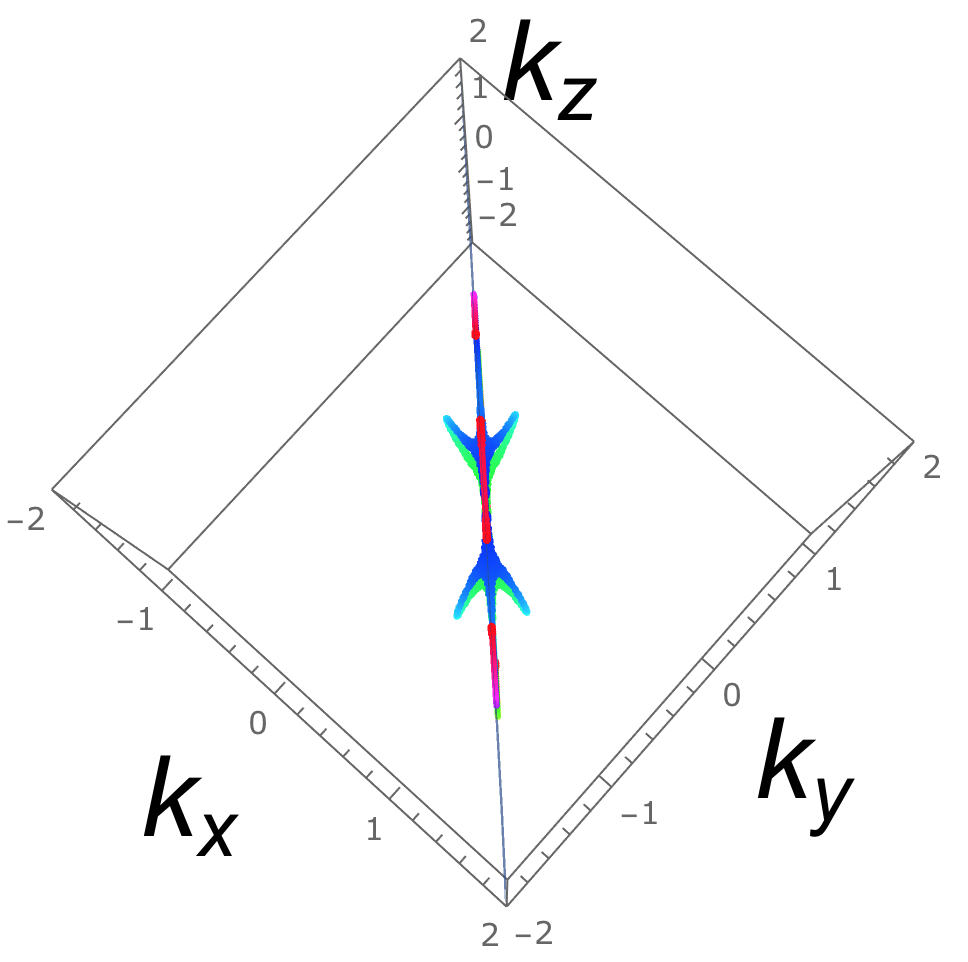}}}
			\end{picture}
	}\\\hrule\quad \\
	\parbox{4.7cm}{(a2)}
	\parbox{4.8cm}{(b2)}
	\parbox{7cm}{(c2)}\\
	
	\parbox{4.7cm}{
		\includegraphics[width=3.5cm]{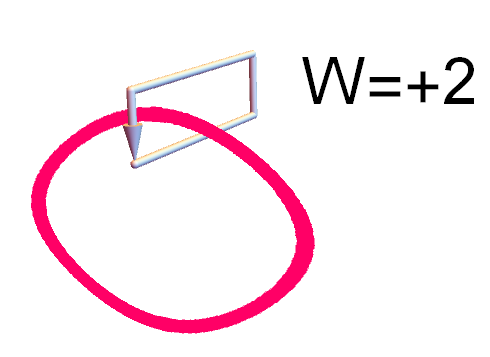}}
	\parbox{4.8cm}{
		\includegraphics[width=4.8cm]{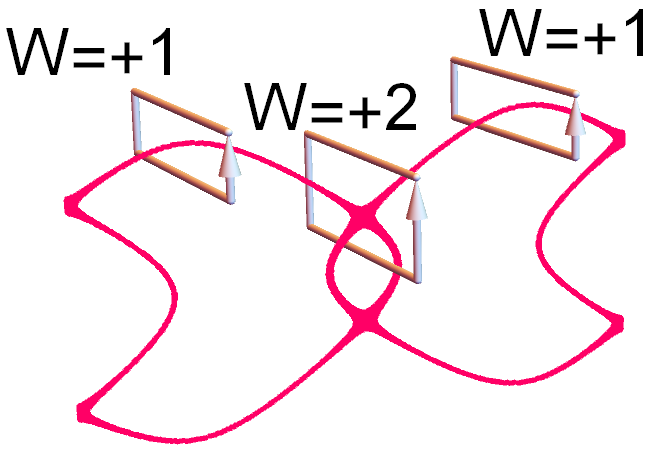}}
	\parbox{7cm}{
		\includegraphics[width=5cm]{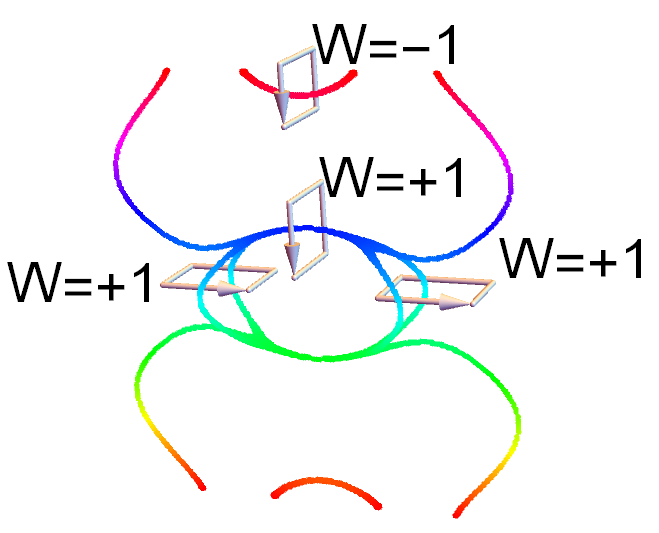}}\\
	\hrule
		\caption{\label{fig:fs} Nodal line Fermi surfaces for various states in Table \ref{tab:U1kitaev}. (a1)-(a2) Typical Fermi surface for states \#1, \#2, \#3, and \#7, shown by the pink loop. The figure corresponds to state \#1 with $ s=t_x=1 $. The winding number is $ +2 $ because there are two degenerate nodal rings. (b1-b2) Typical Fermi surface for the state \#8, shown by the connected pink loops. The figure uses parameters $ t_x=\tilde{t}_x=\tilde{t}_y=\tilde{t}_z=1 $. Each ring has winding number $ +1 $. In the overlapping region the winding number adds up. (c1-c2) One possible stable Fermi surface for states \#4, \#5 and \#6, shown by the multiple connected loops. The figure corresponds to state \#4 with parameters $ (s,t_x,\tilde{t}_x, \tilde{t}_y, \tilde{t}_z) = (0.8, -0.5, 0.7, 0.4, 0.7) $. The two rings in the middle are out of the plane $ k_x+k_y=0 $ while the other parts are in the plane. The inset shows the top view. (b2) shows the winding number for each ring.
	}
	\end{figure*}

For a spinon mean-field Hamiltonian with the form of Eq. (\ref{hmfD}), its structure of spinon fermi surface is determined by the zero mode condition $ |\det(D_{\boldsymbol{k}})|=0 $, which leads to 2 real equations for 3 variables $ (k_x,k_y,k_z) $ in the Brillouin zone. Therefore a one-dimensional fermi surface, i.e. a nodal line, is expected to exist within certain parameter range. The 8 $U(1)$ states (\ref{hcase1})--(\ref{hcase8}) exhibit three typical structures of nodal rings, all of which are topologically stable. We show these nodal line Fermi surfaces in Fig.~\ref{fig:fs} and describe their characters below.
	\begin{itemize}
		\item For states \#1, \#2, \#3 and \#7, the Hamiltonian is labeled by parameters $ \{s, \tilde{s}, t^x \} $. The nodal Fermi surface condition is reduced to a simple form
		\begin{align}
		\nonumber
		&k_x+k_y=0,\quad \cos(k_x-k_y)+\cos(2k_z)=f(s,\tilde{s},t^x)
		,\\
		&f(s,\tilde{s},t^x) = \left\{
		\begin{array}{ll}
		(t^x/\tilde{s})^2, & \mbox{state \#1, \#7},\\
		(s^2+2(\tilde{t}^x)^2)/2\tilde{s}^2, & \mbox{state \#2, \#3}.
		\end{array}
		\right.
		\end{align}
		Thus, for $ |f(s,\tilde{s},t^x)|<2 $, the fermi surface consists of one single nodal ring, which is the intersection of the plane $ k_x+k_y=0 $ and the tube $ \cos(k_x-k_y)+\cos(2k_z)=f $, see Fig.~\ref{fig:fs}(a1). In this case, the nodal ring is the same as the zero-flux Kitaev model discussed in Ref.\onlinecite{Schaffer2015}.
		
		\item For state \#8, the 4 parameters $ \{ t^x, \tilde{t}^x,\tilde{t}^y, \tilde{t}^z \} $ could give rise to two nodal rings in a plane:
		\begin{align}\nonumber
		&k_x+k_y=0,\\ \nonumber
		&(t^x)^4+4A^2+4A(-(t^x)^2+B\cos 2k_z)\cos 2k_y \\
		&+ B(-2(t^x)^2\cos k_z + A(\cos 4k_y + \cos 4k_z))=0,
		\end{align}
		where $ A=-\tilde{t}^x \tilde{t}^y + (\tilde{t}^z)^2$, $B=(\tilde{t}^x)^2 + (\tilde{t}^y)^2 + (\tilde{t}^z)^2 $.
		There are parameter regions where the two rings overlap with each other forming a coplanar network of nodal rings, see Fig.~\ref{fig:fs}(b1) for illustration.
		
		\item For states \#4, \#5, \#6, the 5 parameters $ \{ s,t^x, \tilde{t}^x, \tilde{t}^y, \tilde{t}^z \} $ could give rise to 3 sets of rings. Two of them are in the plane $ k_x+k_y=0 $, while one of them are out of the plane, see Fig.~\ref{fig:fs}(c1). Note that the rings in the middle are connected to the rings above and below them within a finite parameter range, forming a stable three-dimensional network of nodal Fermi surfaces.
		
	\end{itemize}
	
	The three types of spinon nodal rings discussed above turn out to be all topologically stable. Due to the sublattice/chiral symmetry (\ref{chiral sym}),  the mean field ans\"{a}tze (\ref{hmfD}) belongs to the symmetry class AIII. Class AIII supports robust nodal line fermi surfaces\cite{Matsuura2013}, whose topological invariant is given by the following winding number\cite{Schaffer2015}
	\begin{equation}
	W=\frac{1}{4\pi i}\oint d\boldsymbol{k} \cdot \mathrm{Tr}\left(
	D_{\boldsymbol{k}}^{-1} \vec{\nabla}_{\boldsymbol{k}} D_{\boldsymbol{k}} - (D^\dagger_{\boldsymbol{k}})^{-1} \vec{\nabla}_{\boldsymbol{k}} D^\dagger_{\boldsymbol{k}}
	\right),
	\end{equation}
	where $ D_{\boldsymbol{k}} $ is given by Eq.~(\ref{dmat}), and the line integral is done along a loop encircling the nodal ring. The loops in various cases are illustrated in Fig.~\ref{fig:fs}(a2)--(c2).
	
	For the case of a single nodal ring (Fig.~\ref{fig:fs}(a2)), the winding number turns out to be 2, in contrast to a unit winding number in the hyperhoneycomb Kitaev spin liquid\cite{Schaffer2015}. This is due to an accidental two-fold degeneracy on the nodal rings in these $U(1)$ spin liquids, which is an artifact of the 4-band NN model. Each of the two degenerate nodal rings contribute a winding number 1, hence a total winding number 2. Such an accidental degeneracy can be lifted by general spinon hoppings beyond NNs, and is already absent in the case of Fig.~\ref{fig:fs}(b2) where we obtain the unit winding number for each of the linked topological nodal rings.
	
	In the case of three-dimensional linked nodal rings in Fig.~\ref{fig:fs}(c2), we see that all of the nodal line Fermi surfaces are topologically protected with non-zero winding numbers. The nesting of nodal rings is not a result of fine-tuning and turns out to be stable in a finite parameter range. For instance, in the case of Fig.~\ref{fig:fs}(c2), upon decreasing the parameter $ s=0.8 $, the middle two rings will shrink and separate further away from each other, but they are still connected with the upper/lower rings. They finally shrink to two points around $ s\approx 0.3 $, and the upper/lower coplanar rings touch each other. In addition to changing $s$, we have also varied all other parameters and verified the stability of these nodal ring networks.

	\subsection{Discussions}
	The nodal line Fermi surfaces result from any mean-field spinon Hamiltonian of the form (\ref{hmfD}), where the sublattice (or chiral) symmetry (\ref{chiral sym}) is preserved. Due to the sublattice symmetry, the zero mode condition of the off-diagonal Hamiltonian is reduced to $ |\det D_{\boldsymbol{k}}|=0 $, which typically gives rise to the nodal line structure. Sublattice symmetry also excludes all possible mass terms that can gap out the nodal line Fermi surface, hence protecting its topological stability.

On the bipartite hyperhoneycomb lattice, the nodal line Fermi surfaces are not limited to the solvable Kitaev model, or the 8 root $U(1)$ spin liquids in proximity to the Kitaev spin liquid. In fact, only the projective time-reversal symmetry with $G_{\mathcal{T}}(s=1,3)=\tau_0$ and $G_{\mathcal{T}}(s=2,4)=-\tau_0$ is necessary to protect the sublattice symmetry and to host the nodal line Fermi surface. As shown in Appendix \ref{app:u1result_spatial} and \ref{app:z2result}, there are 30 symmetric $U(1)$ spin liquids and 16 symmetric $Z_2$ spin liquids, whose NN mean-field ans\"{a}tze can all support robust nodal line Fermi surfaces.

	\section{Conclusion}

In this work, we classified symmetric quantum spin liquids on the hyperhoneycomb lattice and studied their physical properties. Within the Abrikosov-fermion representation, we obtained 176 $U(1)$ spin liquid states and 160 $Z_2$ states, many of which feature nodal-ring-shaped spinon Fermi surfaces, which are protected by a sublattice symmetry in their nearest-neighbor mean-field ans\"{a}tz. In three dimensions, $U(1)$ spin liquids do not have a finite temperature transition while the $Z_2$ spin liquids would exhibit a thermal transition. Hence future specific heat measurement on $\beta$-Li$_2$IrO$ _3$ under pressure will provide a useful guide for further theoretical investigation of quantum spin liquid phases in this material. We show that 8 ``root'' U(1) spin liquid phases are in proximity to the solvable Kitaev spin liquid. The nodal-line spectra in
these spin liquid phases would give rise to the pseudo gap or the power-law gap in the specific heat coefficient $C/T$ and thermal conductivity $\kappa/T$.
This special set of spin liquid phases may be promising candidates for for the pressurized hyperhoneycomb iridate $ \beta $-Li$_2$IrO$ _3 $ 
given the significant presence of the Kitaev interaction in the parent material. Now that the microscopic spin model for $\beta$-Li$_2$IrO$ _3 $ has been inferred from ab initio calculations at various pressures\cite{Kim2016, Yadav2018}, our results pave the way for future variational Monte Carlo studies on the energetics of these candidate states, which will shed light on the nature of the high-pressure paramagnetic ground state observed in $\beta$-Li$_2$IrO$ _3 $.

	\section{Acknowledgement}
	This work is supported by the U.S. ARO (W911NF-11-1-0230), AFOSR (FA9550-16-1-0006), MURI-ARO (W911NF-17-1-0323) (BH), and National Science Foundation under award number DMR-1653769 (YML), and the NSERC of Canada and the Center for Quantum Materials at the University of Toronto (WC and YBK).
	
	\appendix
	
	\section{Solutions of algebraic PSG equations}\label{app:psg}
	In this section, we present details of calculations to classify all gauge inequivalent PSG's.
	
	\subsection{Space group symmetry}
	The hyperhoneycomb lattice possesses the space group symmetry of Fddd. A general symmetry operation can be written as
	\begin{equation}
	U={\cal T}^{\nu_{\cal T}} T_1^{\nu_1} T_2^{\nu_2} T_3^{\nu_3} \sigma^{\nu_\sigma} r_1^{\nu_{r_1}} r_2^{\nu_{r_2}},
	\end{equation}
	where $ \nu_{1,2,3}\in \mathbb{Z} $ and $ \nu_{\cal T}, \nu_\sigma, \nu_{r_1}, \nu_{r_2}\in\mathbb{Z}_2 $. It consists of time-reversal $ \cal T $, three translations $ T_1, T_2, T_3 $, one inversion $ \sigma $ with respect to the axis origin, and two glide reflections $ r_1, r_2 $. The commutation relations among these symmetry operations imply
	\begin{align}
	&\begin{aligned}
	&T_1^{-1}r_1^{-1}T_1r_1=\boldsymbol{e},\\
	&T_3^{-1}T_1 r_1^{-1} T_2^{-1}r_1=\boldsymbol{e},
	\end{aligned}
	\qquad
	\begin{aligned}
	&T_2^{-1}T_1r_1^{-1} T_3^{-1}r_1 = \boldsymbol{e}, \\
	&T_1^{-1}r_1^2 = \boldsymbol{e},
	\end{aligned}\\ \nonumber \\
	&\begin{aligned}
	&T_1^{-1}T_2 r_2^{-1}T_3^{-1}r_2 = \boldsymbol{e}, \\
	&T_3^{-1}T_2r_2^{-1}T_1^{-1}r_2=\boldsymbol{e},
	\end{aligned}
	\qquad
	\begin{aligned}
	&T_2^{-1}r_2^{-1}T_2r_2=\boldsymbol{e},\\
	&T_2^{-1}r_2^2 = \boldsymbol{e},
	\end{aligned}\\ \nonumber \\
	&\begin{aligned}
	&T_\mu^{-1} \sigma^{-1} T_\mu^{-1} \sigma = \boldsymbol{e},
	\end{aligned}
	~\qquad
	\begin{aligned}
	&\sigma^2=\boldsymbol{e},
	\end{aligned}\\
	& (\sigma r_1)^2 = \boldsymbol{e},~~
	(\sigma r_2)^2 = \boldsymbol{e},~~
	T_3^{-1}r_2r_1^{-1}r_2^{-1}r_1 = \boldsymbol{e}.
	\end{align}

	%The commutation relations for these operations are
	%\begin{align}
	%&
	%r_1T_1=T_1r_1, \quad
	%r_1T_2=T_1T_3^{-1}r_1,\quad
	%r_1T_3=T_1T_2^{-1}r_1, \nonumber \\
	%&
	%r_1^2=T_1\\
	%&
	%r_2T_1=T_2T_3^{-1}r_2,\quad
	%r_2T_2=T_2r_2,\quad
	%r_2T_3=T_1^{-1}T_2r_2, \nonumber \\
	%&
	%r_2^2=T_2\\
	%&\sigma T_\mu = T_\mu^{-1}\sigma, \quad \sigma^2=\boldsymbol{e},\\
	%&
	%r_1\sigma = T_1\sigma r_1,\quad
	%r_2\sigma = T_2\sigma r_2,\quad
	%r_2r_1= T_1^{-1}T_2 r_1r_2.
	%\end{align}
	Due to the nonsymmorphic nature of the glide reflections, the commutations among translations
	\begin{equation}\label{translationcomm}
	[T_\mu, T_\nu]=0
	\end{equation}
	are not independent relations, i.e., sandwiching the following equations
	\begin{multline}
	\nonumber
	T_1T_2T_1^{-1}T_2^{-1} = T_1 r_2^2 (r_2^2 T_1)^{-1} = T_1 r_2^2 (r_2 T_2 T_3^{-1}r_2)^{-1} \\
	=T_1r_2T_3 T_2^{-1}r_2^{-1} = T_1T_1^{-1}T_2r_2T_2^{-1}r_2^{-1} = \boldsymbol{e}
	\end{multline}
	with $ r_1 (\dots) r_1^{-1} $ and $ r_2(\dots) r_2^{-1} $ gives (\ref{translationcomm}).
			
	\subsection{PSG solutions for the U(1) spin liquids}
	\subsubsection{Projective symmetry group constraints}
	
	The projective symmetry group is an extension of the symmetry group to accompany each symmetry operation $U$ with a gauge transformation $G_U$,
	\begin{multline}
	G_U U =(G_{\cal T}{\cal T})^{\nu_{\cal T}} (G_{T_1}T_1)^{\nu_1} (G_{T_2}T_2)^{\nu_2} (G_{T_3}T_3)^{\nu_3} \\
	\cdot (G_\sigma \sigma)^{\nu_\sigma} (G_{r_1}r_1)^{\nu_{r_1}} (G_{r_2}r_2)^{\nu_{r_2}}.
	\end{multline}
	The commutation relations read
	\begin{flalign}
	&\quad G_1^{-1}(r_1(i)) G_{r_1}^{-1}(T_1(i)) G_1(T_1(i)) G_{r_1}(i) = g_3(\theta_{r_11}), & \nonumber \\
	&\quad G_2^{-1}(r_1(i)) G_1(r_1(i)) G_{r_1}^{-1}(i) G_3^{-1}(T_3(i)) & \nonumber  \\
	&\quad  \qquad\qquad\qquad\quad~ \cdot G_{r_1}(T_3(i)) = \tau_0, \quad \text{(IGG of $ G_3 $)} & \nonumber \\
	&\quad G_3^{-1}(r_1(i)) G_1(r_1(i)) G_{r_1}^{-1}(i) G_2^{-1}(T_2(i)) & \nonumber \\
	&\quad  \qquad\qquad\qquad\qquad\qquad\quad~ \cdot G_{r_1}(T_2(i)) = g_3(\theta_{r_13}), & \nonumber \\
	&\quad G_1^{-1}(r_1(i)) G_{r_1}(r_1(i)) G_{r_1}(i) = \tau_0,~~\quad \text{(IGG of $ G_1 $)} &
	\label{set1}
	\end{flalign}
	\begin{flalign}
	&\quad G_1^{-1}(r_2(i)) G_2(r_2(i)) G_{r_2}^{-1}(i) G_3^{-1}(T_3(i)) & \nonumber \\
	&\quad  \qquad\qquad\qquad\qquad\qquad\quad \cdot G_{r_2}(T_3(i)) = g_3(\theta_{r_21}), & \nonumber \\
	&\quad G_2^{-1}(r_2(i)) G_{r_2}^{-1}(T_2(i))G_2(T_2(i)) G_{r_2}(i) = g_3(\theta_{r_2 2}), & \nonumber \\ 
	&\quad G_3^{-1}(r)2(i)) G_2(r_2(i)) G_{r_2}^{-1}(i) G_1^{-1}(T_1(i)) & \nonumber \\
	&\quad \qquad\qquad\qquad\qquad\qquad\quad \cdot G_{r_2}(T_1(i)) = g_3(\theta_{r_23}), & \nonumber \\
	&\quad G_{2}^{-1}(r_2(i)) G_{r_2}(r_2(i)) G_{r_2}(i) = \tau_0,\quad ~ \text{(IGG of $ G_2 $)} &
	\label{set2}
	\end{flalign}
	\begin{flalign}
	&\quad G_1^{-1}(\sigma(i)) G_{\sigma}^{-1}(i) G_1^{-1}(T_1(i)) G_{\sigma}(T_1(i)) = g_3(\theta_{\sigma 1}),& \nonumber \\ 
	&\quad G_2^{-1}(\sigma(i)) G_{\sigma}^{-1}(i) G_2^{-1}(T_2(i)) G_{\sigma}(T_2(i)) = g_3(\theta_{\sigma 2}),& \nonumber \\
	&\quad G_3^{-1}(\sigma(i)) G_{\sigma}^{-1}(i) G_3^{-1}(T_3(i)) G_{\sigma}(T_3(i)) = g_3(\theta_{\sigma 3}),& \nonumber \\
	&\quad G_{\sigma}(\sigma(i)) G_{\sigma}(i) = g_3(\theta_\sigma).&
	\label{set3}
	\end{flalign}
	\begin{flalign}
	&\quad G_\sigma^{-1}(r_1(i)) G_{r_1}(r_1(i)) G_\sigma(i) G_{r_1}(\sigma(i)) = g_3(\theta_{\sigma r_1}), &\nonumber \\
	&\quad G_\sigma^{-1}(r_2(i)) G_{r_2}(r_2(i)) G_\sigma (i) G_{r_2}(\sigma(i)) = g_3(\theta_{\sigma r_2}), &\nonumber \\ 
	&\quad G_3^{-1}(T_3r_1^{-1}(i)) G_{r_2}(T_3r_1^{-1}(i)) G_{r_1}^{-1}(r_2^{-1}(i)) &\nonumber \\
	&\qquad\qquad \qquad\qquad\qquad \cdot G_{r_2}^{-1}(i) G_{r_1}(i) = g_3(\theta_{r_1r_2}).&
	\label{set4}
	\end{flalign}
	Here we have used the invariant gauge group (IGG) to set certain U(1) phases to be zero.
	
	In the canonical gauge, where all fluxes in the ans\"{a}tze point along $ \tau_3 $,
	\begin{equation}
	u_{ij} = u_{ij}^{(0)}\tau_0 + u_{ij}^{(3)}\tau_3 \equiv i\rho_{ij} e^{i\theta_{ij}\tau_3}.
	\end{equation}
	Then PSG constraint $ G_{U}(i) u_{U^{-1}(i), U^{-1}(j)} G_{U}^\dagger(j) = u_{ij}$ implies that the gauge transformations must take one of the following forms at all sites:
	\begin{equation}
	G_U(i) = e^{i\theta_U(i)\tau_3}\equiv g_3(\theta_U(i))
	\end{equation}
	or
	\begin{equation}
	G_U(i) = g_3(\theta_U(i)) (i\tau_1).
	\end{equation}
	
	The nonsymmorphic symmetry constrains the possible form of gauge transformations associated with translations. From $ r_1^2=T_1$ and  $r_2^2=T_2 $,	
	\begin{equation}
	G_1(i), G_2(i) \sim g_3(\theta_{1,2}(i)),
	\end{equation}
	i.e., $G_{1,2}(i)$ cannot have the form $ g_3(\theta_{1,2}(i))(i\tau_1) $ due to the nonsymmorphic glide reflections. Similarly, the constraint $ T_2^{-1}T_1r_1^{-1}T_3^{-1}r_1 = \boldsymbol{e} $ implies
	\begin{equation}
	G_3(i)\sim g_3(\theta_3(i)),
	\end{equation}
	and forbids the form $g_3(\theta_3(i))(i\tau_1)$.
	
	Furthermore, we can use the local gauge freedom to set
	\begin{multline}
	G_1(n_1,n_2,n_3,s) =
	G_2(0,n_2,n_3,s) \\
	= G_3(0,0,n_3,s) = 1.
	\end{multline}	
	Using the constraints from the commutation relations $ T_\mu^{-1}T_\nu^{-1}T_\mu T_\nu = \boldsymbol{e}$,
	\begin{align}\nonumber
	&G_1(n_1,n_2,n_3,s)=1, \\ \nonumber
	&G_2(n_1,n_2,n_3,s)=g_3(n_1\theta_{12}), \\ 
	&G_3(n_1,n_2,n_3,s)=g_3(n_1\theta_{13}+n_2\theta_{23}).
	\label{eq:Tcomm}
	\end{align}
	This is the only possible form for $ G_{1,2,3}(i)$.

	\subsubsection{Solving for equations (\ref{set1})}
	
	Using (\ref{eq:Tcomm}) and (\ref{r1trans}), we can write (\ref{set1}) as
	\begin{align}\nonumber
	&G_{r_1}(T_1(i)) = g_3(\theta_{r_11}) G_{r_1}(i),\\ \nonumber
	&g_3(-(n_1+n_2+n_3+A_s)\theta_{12}) G_{r_1}^{-1}(i) \\ \nonumber
	&\quad\qquad\qquad\quad \cdot g_3(-n_1\theta_{13}-n_2\theta_{23}) G_{r_1}(T_3(i)) = \tau_0,\\ \nonumber
	&g_3(-(n_1+n_2+n_3+A_s)\theta_{13}+n_3\theta_{23}) G_{r_1}^{-1}(i) \\\nonumber
	&\quad\qquad\qquad\qquad \cdot g_3(-n_1\theta_{12}) G_{r_1}(T_2(i)) = g_3(\theta_{r_13}),\\
	&G_{r_1}(n_1+n_2+n_3+A_s, -n_3, -n_2, r_1(s)) G_{r_1}(i) = \tau_0.
	\label{constrR1}
	\end{align}
	
	\textbf{1)} If $G_{r_1}\sim g_3$: with $ G_{r_1}(0,0,0,s)=g_3(\varphi_{r_1}(s))$,
	\begin{align}\nonumber
	&G_{r_1}(n_1,n_2,n_3,s)= g_3\left(\varphi_{r_1}(s) +n_1\theta_{r_11} + n_2\theta_{r_13} \right) \\ \nonumber
	&\cdot g_3\left( \left[ n_3(n_1+n_2+A_s)+n_1n_2+ \frac{n_3(n_3-1)}{2} \right ]\theta_{12} \right) \\
	&\cdot g_3 \left( \left[n_2(n_1+n_3+A_s)+n_1n_3 + \frac{n_2(n_2-1)}{2} \right ]\theta_{13} \right).
	\end{align}
	Here we used
	\begin{align} \nonumber
	&f_{n+1}= \exp(\alpha n + \beta) f_n \\
	&\Rightarrow f_n= \exp\left(\alpha \frac{n(n-1)}{2} + n\beta \right) f_0.
	\end{align}
		
	From the constraint (\ref{constrR1}),
	\begin{align}
	\nonumber
	&g_3\left(\varphi_{r_1}(r_1(s)) + \varphi_{r_1}(s) + (2n_1+n_2+n_3+A_s)\theta_{r_11} \right) \\ \nonumber
	&\cdot g_3\left((n_2-n_3)\theta_{r_13} -(\theta_{12}+\theta_{13})\left(\frac{n_2(n_2+1)}{2}\right) \right) \\
	&\cdot g_3 \left( -(\theta_{12}+\theta_{13})\left(\frac{n_3(n_3+1)}{2}\right)\right)= \tau_0 \label{eq:const_r11}
	\end{align}
	with
	\begin{equation}
	A_s + A_{r_1(s)} = 1 = A_s + A_{r_2(s)};
	\end{equation}
	$ A_s $ is defined in (\ref{r1trans})-(\ref{r2trans}).
		
	At $ n_i=0 $, we have $ \varphi_{r_1}(3)+\varphi_{r_1}(1) = \varphi_{r_1}(1)+\varphi_{r_1}(3)+\theta_{r_11}
	= 2\pi \mathbb{Z} $. So $\theta_{r_11}=0$. Similarly, $ \theta_{r_13}=0$ because $ \varphi_{r_1}(3)+
	\varphi_{r_1}(1) + \theta_{r_13} = 2\pi\mathbb{Z}$ when $n_2=0$ and $n_3=-1$.
	Since Eq. (\ref{eq:const_r11}) should hold for all sites, $ \theta_{12}=-\theta_{13} $. To sum up,
	\begin{align}\nonumber
	&G_{r_1}(n_1,n_2,n_3,s) = g_3\left(\varphi_{r_1}(s)+(n_3-n_2)A_s\theta_{12} \right) \\
	&\cdot g_3\left( \left[ \frac{n_3(n_3-1)-n_2(n_2-1)}{2} \right]\theta_{12}\right),
	\end{align}
	where $\varphi_{r_1}(1)=-\varphi_{r_1}(3)$ and $\varphi_{r_1}(2) = -\varphi_{r_1}(4).$
	\\
		
	\textbf{2)} If $ G_{r_1}\sim g_3(i\tau_1)$:
	\begin{align}\nonumber
	&G_{r_1}(n_1,n_2,n_3,s)= g_3\left( \varphi_{r_1}(s)+n_1\theta_{r_1 1}+n_2\theta_{r_1 3} \right)\\ \nonumber
	&\cdot g_3 \left(\theta_{12}\left[n_1 n_2-n_3(n_1+n_2+A_s)-\frac{n_3(n_3-1)}{2}\right] \right) \\ \nonumber
	&\cdot g_3 \left(\theta_{13}\left[n_1 n_3-n_2(n_1+n_3+A_s)-\frac{n_2(n_2-1)}{2}\right] \right) \\
	&\cdot g_3 \left(2n_2 n_3\theta_{2 3} \right) (i\tau_1),
	\end{align}
	where $G_{r_1}(0,0,0,s)=g_3(\varphi_{r_1}(s)) (i\tau_1)$.
	Then, the constraint $ G_{r_1}(r_1(i)) G_{r_1}(i) = \tau_0 $ becomes
	\begin{align}
	\nonumber
	g_3 \left(\varphi_{r_1}(r_1(s))-\varphi_{r_1}(s)+A_s\theta_{r_1 1} + (n_2+n_3)(\theta_{r_1 1}-\theta_{r_1 3}) \right)\\
	\cdot g_3\left( \left[ \frac{n_2(n_2+1)}{2} - \frac{n_3(n_3+1)}{2} \right] \left(\theta_{12}-\theta_{13} \right) \right) = \tau_0.
	\end{align}
	Thus, we have $ \varphi_{r_1}(3)-\varphi_{r_1}(1)=0$ and $\varphi_{r_1}(1)-\varphi_{r_1}(3)=\theta_{r_1 1} $ implying
	\begin{enumerate}
	\item $\theta_{r_1 1}=\theta_{r_1 3}=0, \quad \theta_{12}=\theta_{13}$,
	\item $\varphi_{r_1}(1)=\varphi_{r_1}(3),\quad \varphi_{r_1}(2)=\varphi_{r_1}(4)$.
	\end{enumerate}
	Therefore
	 \begin{align}
	&G_{r_1}(n_1,n_2,n_3,s) = g_3\left(\varphi_{r_1}(s)-\theta_{12}\left(2n_2 n_3+(n_2+n_3)A_s \right) \right)\nonumber \\
	&\cdot g_3 \left(-\theta_{12} \left[\frac{n_2(n_2-1)}{2}+\frac{n_3(n_3-1)}{2} \right] + 2n_2 n_3\theta_{23}\right)(i\tau_1).
	\end{align}
	
	\subsubsection{Solving for equations (\ref{set2})}
	
	Using (\ref{eq:Tcomm}) and (\ref{r2trans}), we can rewrite (\ref{set2}) as
	\begin{align}
	&g_3(-n_3 \theta_{12}) G_{r_2}^{-1}(i)g_3(-n_1\theta_{13}-n_2\theta_{23}) \nonumber \\
	&~\qquad\qquad\qquad\qquad\qquad \cdot G_{r_2}(T_3(i)) = g_3(\theta_{r_2 1}),\nonumber \\
	& g_3(n_3 \theta_{12})G_{r_2}^{-1}(T_2(i)) g_3(n_1 \theta_{12})G_{r_2}(i) = g_3(\theta_{r_2 2}), \nonumber \\
	& g_3(n_3 \theta_{13}-(n_1+n_2+n_3+A_s)\theta_{23})g_3(-n_3\theta_{12}) \nonumber \\ 
	&~~\quad\qquad\qquad\qquad \cdot G_{r_2}^{-1}(i) G_{r_2}(T_1(i)) = g_3(\theta_{r_2 3}),\nonumber \\ 
	& g_3(n_3 \theta_{12})G_{r_2}(-n_3,n_1+n_2+n_3+A_s,-n_1,r_2(s)) \nonumber \\
	&~\qquad\qquad\qquad\qquad\qquad \cdot G_{r_2}(n_1,n_2,n_3,s) = \tau_0. \label{eq:r1}
	\end{align}
	
		\textbf{1)} If $ G_{r_2}\sim g_3 $:
		\begin{align}
		&G_{r_2}(n_1,n_2,n_3,s) = g_3\left(\varphi_{r_2}(s) +n_1\theta_{r_2 3} + n_2\theta_{r_2 2} + n_3\theta_{r_2 1} \right) \nonumber \\
		&\cdot g_3 \left( \theta_{12}(n_1 n_2 + n_2 n_3 + n_1 n_3) + \theta_{12}\frac{n_3(n_3-1)}{2}\right) \nonumber \\
		&\cdot g_3 \left( \theta_{23}(n_2 n_3+n_1(n_2+n_3+A_s))+\theta_{23} \frac{n_1(n_1-1)}{2} \right).
		\end{align}
		From the last equation of Eq.~(\ref{eq:r1}),
		\begin{align}
		&g_3 \left( \varphi_{r_2}(r_2(s))+\varphi_{r_2}(s) + (n_1-n_3)(\theta_{r_2 3}-\theta_{r_2 1}) \right) \nonumber \\
		&\cdot g_3 \left( (n_1+2n_2+n_3+A_s) \theta_{r_2 2} + \theta_{12}(n_3-(n_1+n_3)A_s) \right)  \nonumber \\ 
		&\cdot g_3 \left(- \theta_{12}\left[\frac{n_1(n_1-1)}{2} + \frac{n_3(n_3+1)}{2}\right]\right) \nonumber \\
		&\cdot g_3 \left( - \theta_{23}\left[\frac{n_1(n_1+1)}{2}+\frac{n_3(n_3+1)}{2}\right] \right) = \tau_0.
		\end{align}
		First, take $ n_1=n_2=n_3=0$. Then $ \varphi_{r_2}(3)+\varphi_{r_2}(1)=0$ and $\varphi_{r_2}(1)+\varphi_{r_2}(3) = \theta_{r_2 2} $. So $ \theta_{r_2 2}=0 $.
		Next, take $ n_3=0$ and $n_1=1 $. For $ s=1,2 $, $ \theta_{r_2 3}-\theta_{r_2 1}-\theta_{23}=0 $; for $ s=3,4 $, $ \theta_{r_2 3}-\theta_{r_2 1}-\theta_{12}-\theta_{23} = 0 $. Thus, $ \theta_{12}=0 $.
		Finally, when $n_3=0 $, we have $ n_1((\theta_{r_2 3}-\theta_{r_2 1})-(n_1+1)\theta_{23}/2) = 0 $. So $ \theta_{23}=0$ and $\theta_{r_2 3}=\theta_{r_2 1}$.
		\\
		
		\textbf{2)} If $ G_{r_2}\sim g_3 (i\tau_1) $:
		\begin{align}
		& G_{r_2}(n_1,n_2,n_3,s) = g_3 \left ( \varphi_{r_2}(s)+n_1 \theta_{r_2 3}+n_2\theta_{r_2 2}+n_3\theta_{r_2 1} \right) \nonumber\\
		&\cdot g_3\left(\theta_{12}\left[n_1 n_2 - n_2 n_3 - n_1 n_3 - \frac{n_3(n_3-1)}{2}\right] \right) \nonumber \\
		&\cdot g_3 \left(\theta_{23}\left[n_2 n_3 -n_1(n_2+n_3+A_s)-\frac{n_1(n_1-1)}{2}\right] \right) \nonumber \\
		&\cdot g_3 \left( \theta_{13}(2n_1 n_3) \right)(i\tau_1)
		\end{align}
		with
		\begin{align}
		&g_3\left( \varphi_{r_2}(r_2(s))-\varphi_{r_2}(s) + n_1(-\theta_{r_2 3}-\theta_{r_2 1}+\theta_{r_2 2}+A_s\theta_{12})\right) \nonumber \\
		&\cdot g_3 \left(n_3(-\theta_{r_2 3}-\theta_{r_2 1}+\theta_{r_2 2}+(1-A_s)\theta_{12}+\theta_{23}) \right) \nonumber \\
		&\cdot g_3 \left( A_{s}\theta_{r_2 2} \right) = \tau_0.
		\end{align}
		Let $ n_1=n_3=0 $, Then from $ \varphi_{r_2}(3)-\varphi_{r_2}(1)=0$ and $\varphi_{r_2}(1)-\varphi_{r_2}(3)-\theta_{r_2 2}=0 $, we have $ \theta_{r_2 2}=0 $. If we take $ n_3=0, n_1=1 $, we have $ (\theta_{r_2 3}+\theta_{r_2 1})-A_s\theta_{12}+\theta_{23}=0 $. Comparing the equation for $ s=1,2 $ and $ s=3,4 $ gives $ \theta_{12}=0 $. When $ n_3=0 $, we have $n_1(\theta_{r_2 3}+\theta_{r_2 1}+n_1(n_1+1)\theta_{23}/2)=0 $. Hence, $ \theta_{23}=0$ and $\theta_{r_2 1}=-\theta_{r_2 3} $.
		
	In sum,  we found that
	\begin{equation}
	\theta_{12} = \theta_{13} = \theta_{23} = \theta_{r_1 1} = \theta_{r_1 3} = \theta_{r_2 2}=0.
	\end{equation}
	Therefore the gauge transformations associated to the translations are all trivial, i.e.,
	\begin{equation}
	G_1(i)=G_2(i)=G_{3}(i)=\tau_0.
	\end{equation}
	
	Before we proceed, let us use the sublattice dependent local U(1) gauge transformations to gauge away $\varphi_{r_1}(s)$. With $W_{s=1} = g_3(-\varphi_{r_1}(1))$, $W_{s=2}=g_3(-\varphi_{r_1}(2))$, and $W_{s=3}=W_{s=4}=\tau_0$, we can set
	\begin{equation}
	\varphi_{r_1}(1) = \varphi_{r_1}(2)=0,
	\end{equation}
	e.g., $G_{r_1}(s=1) \rightarrow W_{s=1}G_{r_1}(s=1)W_{s=3}^\dagger = \tau_0$. Then the solutions for $ G_{r_1}$ and $G_{r_2} $ can be summarized as
	\begin{align}\nonumber
	& G_{r_1}(i)=\tau_0\quad \mbox{or}\quad G_{r_1}(i)=i\tau_1,\quad\\\nonumber
	& G_{r_2}(n_1,n_2,n_3,s) = g_3(\varphi_{r_2}(s)+(n_1+n_3)\theta_{r_2 3}),
	\\ \nonumber
	& \qquad\varphi_{r_2}(1)=-\varphi_{r_2}(3), \quad \varphi_{r_2}(2)=-\varphi_{r_2}(4)\\ \nonumber
	& \mbox{or}
	\\ \nonumber
	&G_{r_2}(n_1,n_2,n_3,s)=g_3(\varphi_{r_2}(s)+(n_1-n_3)\theta_{r_2 3})(i\tau_1),\\ \label{tempSol4}
	& \qquad \varphi_{r_2}(1)=\varphi_{r_2}(3), \quad \varphi_{r_2}(2)=\varphi_{r_2}(4).
	\end{align}

	\subsubsection{Solving for equations (\ref{set3})}
	Since the translations have trivial gauge groups, we immediately have two cases.
	\\
	
	\textbf{1)} If $ G_{\sigma}\sim g_3 $,
		\begin{align}\nonumber
		&G_\sigma(n_1,n_2,n_3,s) = g_3(\varphi_{\sigma}(s)+n_1\theta_{\sigma 1}+n_2 \theta_{\sigma 2} + n_3\theta_{\sigma 3}), \\ \label{tempSol1}
		&\varphi_\sigma(1)=-\varphi_\sigma(4), \quad \varphi_\sigma(2)=-\varphi_{\sigma}(3).
		\end{align}
		The IGG of $ G_\sigma $ has been used to set $ \theta_\sigma=0 $.
		\\
		
		\textbf{2)} If $ G_\sigma\sim g_3(i\tau_1) $,
		\begin{align}
		\nonumber
		&G_\sigma(n_1,n_2,n_3,s)=g_3(\varphi_\sigma(s)+n_1\theta_{\sigma 1}+n_2\theta_{\sigma 2} + n_3\theta_{\sigma 3})(i\tau_1),\\ \nonumber
		&\theta_{\sigma 1},\theta_{\sigma 2}, \theta_{\sigma 3}, \theta_{\sigma}=0,\pi, \\
		&\varphi_{\sigma}(4)-\varphi_\sigma(1)=\varphi_{\sigma}(3)-\varphi_\sigma(2) = \theta_\sigma.
		\end{align}
		In this case, the IGG of $ G_\sigma $ is unused.
		
		We can again gauge away the $\varphi_\sigma(4)$ using the local $U(1)$ gauge freedom. When $G_\sigma \sim g_3$, $W_{s=1} = W_{s=3} = \tau_0$ and $W_{s=2} = W_{s=4} = g_3(-\varphi_\sigma(4))$ can set
		\begin{equation}\label{tempSolvarphi}
		\varphi_\sigma(4) = 0.
		\end{equation}
		For $G_\sigma \sim g_3(i\tau_1)$, we can do the same gauge fixing with global $U(1)$ gauge rotation $W_s = g_3(-\varphi_\sigma(4)/2)$.
		
	\subsubsection{Solving for equations~(\ref{set4})}
	
	Up to this point, there are $ 2\times2\times2=8 $ classes of solutions due to the choices $ G_{\sigma}, G_{r_1}, G_{r_2}\sim g_3 $ or $ g_3(i\tau_1) $. We solve the 8 classes separately using the solutions (\ref{tempSol4})--(\ref{tempSolvarphi}).
		
	\textbf{1)} $ G_{\sigma}, G_{r_1}, G_{r_2}\sim g_3 $. Then for equations (\ref{set4}), the first equation gives
	\begin{align}
	&g_3\left(-\varphi_\sigma(r_1(s))+\varphi_\sigma(s)+(n_2+n_3)(-\theta_{\sigma 1}+\theta_{\sigma 3}+\theta_{\sigma 2})\right)\nonumber \\
	&\qquad\qquad\qquad \cdot g_3(-A_s\theta_{\sigma 1}))= \tau_0 \qquad\qquad \mbox{(IGG of $ G_{r_1}$)} \nonumber \\
	&\Rightarrow \theta_{\sigma 1}=0, ~ \theta_{\sigma 3}+\theta_{\sigma 2}=0, ~ \varphi_\sigma(s)=0.
	\end{align}
	The second equation gives
	\begin{align}\nonumber
	& g_3\left(\varphi_{r_2}(r_2(s))+\varphi_{r_2}(\sigma(s))+(n_1+n_3)(\theta_{\sigma 3}-\theta_{\sigma 2}-2\theta_{r_2 3}) \right)\\ \nonumber
	&\qquad\qquad\qquad \cdot  g_3( -A_s\theta_{\sigma 2}))=\tau_0 \qquad\qquad \mbox{(IGG of $ G_{r_2}$)}\\ \nonumber
	&\Rightarrow \theta_{\sigma 2}=0, ~ \theta_{r_2 3}=0, \pi,\\
	& \varphi_{r_2}(1)=-\varphi_{r_2}(2)=-\varphi_{r_2}(3)=\varphi_{r_2}(4)\equiv \varphi_{r_2}.
	\end{align}
	The third equation gives
	\begin{align}\nonumber
	& g_3(\varphi_{r_2}(r_1(s))-\varphi_{r_2}(s)+\theta_{r_2 3}A_s) = g_3(\theta_{r_1r_2})\\ \nonumber
	&\Rightarrow ~ \theta_{r_1r_2}=2\varphi_{r_2}=0 \text{ or } \theta_{r_1r_2}=2\varphi_{r_2}=\pi \Rightarrow \theta_{r_2 3}=0
	\end{align}
		In sum, for $ \theta_{r_1r_2}=0 $,
		\begin{equation}
		G_{\sigma}(i)=G_{r_1}(i)=G_{r_2}(i)=1;
		\end{equation}
		for $ \theta_{r_1r_2}=\pi $,
		\begin{align}\nonumber
		&G_{\sigma}(n_1,n_2,n_3,s)=G_{r_1}(n_1,n_2,n_3,s)=1, \\
		&G_{r_2}(n_1,n_2,n_3,s)=\left\{
			\begin{array}{ll}
			i\tau_3, & s=1,4\\
			-i\tau_3, & s=2,3.
			\end{array}
			\right.
		\end{align}
		For the other cases, we can analogously solve the equations in (\ref{set4}) one by one. So we will only briefly sketch the solutions below.
		
		\textbf{2)} If $ G_{\sigma}, G_{r_2}\sim g_3, G_{r_1}\sim g_3(i\tau_1) $, the first equation gives
		\begin{align}\nonumber
		&\theta_{\sigma 1}=0, ~ \theta_{\sigma 2}=\theta_{\sigma 3}, ~ \theta_{\sigma r_1} = \varphi_\sigma = 0, \pi\\
		&\varphi_\sigma(1)+\varphi_\sigma(3) = \theta_{\sigma r_1} = \varphi_\sigma(2)+\varphi_{\sigma}(4).
		\end{align}
		The second equation, with $ \theta_{\sigma r_2}=0 $ set by the IGG of $ G_{r_2} $, gives
		\begin{align}\nonumber
		& \theta_{\sigma 2}=\theta_{\sigma 3}=0, &\varphi&_{\sigma} = 0,\pi\\
		& \varphi_{r_2}(1,3)=0, & \varphi&_{r_2}(2,4)=\varphi_{\sigma}.
		\end{align}
		The last equation gives
		\begin{equation}
		\theta_{r_2 3}=\theta_{r_1r_2} = 0.
		\end{equation}
		
		\textbf{3)} If $ G_{\sigma}, G_{r_1}\sim g_3, G_{r_2}\sim g_3(i\tau_1) $, the first equation, with IGG of $ G_{r_1} $ forcing $ \theta_{\sigma r_1}=0 $, gives
		\begin{align}
		\theta_{\sigma 1}=0, ~ \theta_{\sigma 2}+\theta_{\sigma 3}=0, ~ \varphi_{\sigma}(s)=0.
		\end{align}
		The second equation gives
		\begin{align}\nonumber
		&\theta_{\sigma 2}, \theta_{\sigma 3}=0, \quad \theta_{\sigma r_2}=0, \pi,\\
		&\varphi_{r_2}(1)=\varphi_{r_2}(3)=0, \quad \varphi_{r_2}(2)=\varphi_{r_2}(4)=\theta_{\sigma r_2}.
		\end{align}
		The last equation gives
		\begin{equation}
		\theta_{r_1r_2}=\theta_{r_2 3}=0.
		\end{equation}
		
		\textbf{4)} If $ G_{r_1}, G_{r_2}\sim g_3, G_\sigma \sim g_3(i\tau_1) $, the first equation gives
		\begin{align}\nonumber
		&\theta_{\sigma 1}=0,  &\theta&_{\sigma 2}=-\theta_{\sigma 3}, &\theta&_{\sigma},\theta_{\sigma r_1}=0,\pi \\
		&\varphi_\sigma (1) = \theta_\sigma, &\varphi&_{\sigma}(2)=\theta_{\sigma r_1}, &\varphi&_\sigma(3)=\theta_{\sigma r_1}+\theta_\sigma, \nonumber \\
		&\varphi_\sigma(4)=0. & & & &
		\end{align}
		The second equation, with IGG of $ G_{r_2} $ setting $ \varphi_{r_2}(1)=0 $, gives
		\begin{align}\nonumber
		& \theta_{\sigma 2}=0, &\theta&_{\sigma r_2}=0, \pi,\\
		& \varphi_{r_2}(1,3)=0, &\varphi&_{r_2}(2,4)=\theta_{\sigma r_1}+\theta_{\sigma r_2}
		\end{align}
		The third equation gives
		\begin{equation}
		\theta_{r_1r_2}=\theta_{r_2 3}=0.
		\end{equation}
		
		\textbf{5)} If $ G_\sigma,G_{r_1}\sim g_3(i\tau_1), G_{r_2}\sim g_3 $, the first equation, with the IGG of $ G_{\sigma} $ setting $ \theta_{\sigma r_1}=0 $, gives
		\begin{align}
		&\theta_{\sigma 1}=0, &\varphi&_\sigma=\theta_\sigma =0, \pi, &\theta&_{\sigma 2 }=\theta_{\sigma 3}.
		\end{align}
		The second equation gives
		\begin{align}\nonumber
		&\theta_{\sigma r_2}=0, \pi, &\theta&_{\sigma 2}=0,\\
		&\varphi_{r_2}(1,3)=0, &\varphi&_{r_2}(2,4)=\theta_{\sigma r_2}.
		\end{align}
		Finally, the third equation, with IGG of $ G_{r_2} $ setting $ \theta_{r_1r_2}=0 $, gives
		\begin{eqnarray}
		\theta_{r_2 3}=0.
		\end{eqnarray}

		\textbf{6)} If $ G_\sigma, G_{r_2}\sim g_3(i\tau_1), G_{r_1}\sim g_3 $, the first equation gives
		\begin{align}\nonumber
		&\theta_{\sigma r_1}=0,\pi, &\theta&_{\sigma 1}=0, &\theta&_{\sigma 2}=-\theta_{\sigma 3},\\
		&\varphi_{\sigma}(1)=\theta_\sigma, &\varphi&_{\sigma}(2)=\theta_{\sigma r_1}, &\varphi&_{\sigma}(3)=\theta_{\sigma}+\theta_{\sigma r_1}, \nonumber \\
		&\varphi_{\sigma}(4)=0. & & & &
		\end{align}
		The second equation, with IGG of $ G_\sigma $ setting $ \theta_{\sigma r_2}=0 $ and IGG of $ G_{r_2} $ setting $ \varphi_{r_2}(1)=0 $, gives
		\begin{align}
		&\theta_{\sigma r_1}=0, \pi, &\theta&_{\sigma 2}=0, \nonumber \\
		&\varphi_{r_2}(1,3)=0, &\varphi&_{r_2}(2,4)=\theta_{\sigma r_1}.
		\end{align}
		Finally, the third equation, with IGG of $ G_{r_1} $ setting $ \theta_{r_1r_2}=0 $, gives
		\begin{equation}
		\theta_{r_2 3}=0.
		\end{equation}
		
		\textbf{7)} If $ G_{r_1}, G_{r_2}\sim g_3(i\tau_1), G_\sigma \sim g_3 $, the first equation gives
		\begin{align}\nonumber
		&\theta_{\sigma r_1}=0, \pi, &\theta&_{\sigma 1}=0, &\theta&_{\sigma 2}=\theta_{\sigma 3}, \nonumber \\
		&\varphi_{\sigma}(1)=\varphi_{\sigma}(4)=0, &\varphi&_{\sigma}(2)=\varphi_{\sigma}(3)=\theta_{\sigma r_1}.
		\end{align}
		The second equation gives
		\begin{flalign}
		&\theta_{\sigma 2}=0, &\varphi&_{r_2}(1,3)=0, &\varphi_{r_2}(2,4)=\theta_{\sigma r_1}+\theta_{\sigma r_2}.&
		\end{flalign}
		The last equation, with IGG of $ G_{r_2} $ giving $ \theta_{r_1r_2}=0 $, provides
		\begin{equation}
		\theta_{r_23}=0.
		\end{equation}
		
		\textbf{8)} If all of $ G_{\sigma}, G_{r_1}, G_{r_2}\sim g_3(i\tau_1) $: the first equation, with IGG of $ G_\sigma $ setting $ \theta_{\sigma r_1}=0 $, gives
		\begin{align}
		&\theta_{\sigma 1}=0,  &\theta&_{\sigma 2}=\theta_{\sigma 3}, \nonumber \\
		&\varphi_\sigma(1,3)=\theta_\sigma, &\varphi&_{\sigma}(2,4)=0.
		\end{align}
		The second equation with IGG of $ G_{r_2} $ setting $ \theta_{\sigma r_2}=0 $, gives
		\begin{flalign}
		&\theta_{\sigma 2}=0, &\varphi&_{r_2}(1)=-\varphi_{r_2}(2), &\varphi_{r_2}(3)=-\varphi_{r_2}(4).
		\end{flalign}
		Finally, the last equation gives
		\begin{eqnarray}
		\theta_{r_1r_2}=0,\pi, \quad \theta_{r_23}=0.
		\end{eqnarray}
		%Need to make connection between \varphi_{r_2}(s) and \theta_{r_1 r_2}.

	\subsection{Summary: U(1) PSG solutions with space group symmetry only}\label{app:u1result_spatial}
	Now we summarize the solutions obtained in the previous section. First, all of the three gauge transformations associated with translations are trivial
	\begin{eqnarray}
	G_1(i)=G_2(i)=G_3(i) = \tau_0,
	\end{eqnarray}
	and $ G_\sigma(i) = g_3(s) (i\tau_1)^{n_\sigma}$, $G_{r_1}(i)=g_3(s)(i\tau_1)^{n_{r_1}} $ and $ G_{r_2}(i) = g_3(s)(i\tau_1)^{n_{r_2}} $ all depend only on the sublattice sites. This is a consequence of the nonsymmorphic glide reflection symmetry.
	
	With all different combination of $ n_\sigma, n_{r_1}, n_{r_2}=\pm1 $, there are 8 classes of solutions for $G_\sigma$, $G_{r_1}$, and $G_{r_2}$. There are total 30 solutions, divided into 8 classes by $ (n_\sigma, n_{r_1},n_{r_2}) $.
	Recall that $ g_3(\theta) \equiv e^{i\theta\tau_3} $. (Note that $ g_3(0)=1$, $g_3(\pi) = -1$, $g_3(\pm\pi/2)=\pm (i\tau_3) $.) 
	\begin{enumerate}
		\item
		Class (0,0,0): $ \theta_{r_1r_2}=0,\pi $, \quad [2 solutions]
		\begin{align}\nonumber
		&G_\sigma = G_{r_1} = \tau_0, \\ \label{classA}
		&G_{r_2} = \left\{ \begin{array}{ll}
		g_3(\theta_{r_1r_2}/2), & s=1,4\\
		g_3(-\theta_{r_1r_2}/2), & s=2,3
		\end{array}\right.	
		\end{align}
		\item
		Class (0,1,0): $ \theta_{\sigma r_1}=0,\pi $, \quad [2 solutions]
		\begin{align}
		\nonumber
		&G_{\sigma} = \left\{ \begin{array}{ll}
		\tau_0, & s=1,4\\
		g_3(\theta_{\sigma r_1}), & s=2,3
		\end{array} \right.,\\ \nonumber
		&G_{r_1} = i\tau_1,\\
		&G_{r_2}= \left\{ \begin{array}{ll}
		\tau_0, & s=1,3\\
		g_3(\theta_{\sigma r_1}), & s=2,4
		\end{array}\right.
		\end{align}
		\item
		Class (0,0,1): $ \theta_{\sigma r_2} = 0,\pi $, \quad [2 solutions]
		\begin{align}\nonumber
		&G_{\sigma}=G_{r_1} = \tau_0, \\
		&G_{r_2} = \left\{
		\begin{array}{ll}
		(i\tau_1), & s=1,3\\
		g_3(\theta_{\sigma r_2})(i\tau_1) , & s=2,4
		\end{array}\right.
		\end{align}
		\item
		Class (1,0,0): $\theta_{\sigma}, \theta_{\sigma r_1}, \theta_{\sigma r_2} = 0,\pi$, \quad [8 solutions]
		\begin{align}
		\nonumber
		&G_\sigma = g_3(\varphi_\sigma(s))(i\tau_1), \quad \varphi_\sigma(1)=\theta_\sigma, \quad\varphi_\sigma (2) = \theta_{\sigma r_1},\\ \nonumber
		&\quad \varphi_\sigma (3) = \theta_{\sigma r_1}+\theta_\sigma , \quad \varphi_\sigma (4) = 0;\\ \nonumber
		&G_{r_1} = \tau_0, \quad G_{r_2} = g_3(\varphi_{r_2}(s)), \quad \varphi_{r_2}(1)=\varphi_{r_2}(3)=0, \\
		&\quad\varphi_{r_2}(2) = \varphi_{r_2}(4) =  \theta_{\sigma r_1} + \theta_{\sigma r_2}.
		\end{align}
		\item
		Class (1,1,0): $ \theta_{\sigma}, \theta_{\sigma r_2} = 0,\pi $, \quad [4 solutions]
		\begin{align}
		\nonumber
		&G_\sigma = g_3(\varphi_\sigma(s)) (i\tau_1), \\ \nonumber
		&\quad
		\varphi_\sigma (1) = \varphi_\sigma(3) = \theta_\sigma, \quad \varphi_\sigma(2) = \varphi_\sigma(4)=0\\ \nonumber
		&G_{r_1}=(i\tau_1), \quad G_{r_2}=g_3(\varphi_{r_2}(s)),\\ 
		&\quad \varphi_{r_2}(1)=\varphi_{r_2}(3) = 0, \quad \varphi_{r_2}(2)=\varphi_{r_2}(4) = \theta_{\sigma r_2}.
		\end{align}
		\item
		Class (1,0,1): $ \theta_\sigma, \theta_{\sigma r_1} = 0,\pi $, \quad [4 solutions]
		\begin{align} \nonumber
		&G_\sigma = g_3(\varphi_\sigma(s))(i\tau_1), \quad
		\varphi_\sigma(1)=\theta_\sigma, \quad \varphi_\sigma(2)=\theta_{\sigma r_1},\\ \nonumber
		&\quad \varphi_\sigma(3)=\theta_\sigma+\theta_{\sigma r_1},\quad \varphi_\sigma(4)=0\\ \nonumber
		&G_{r_1} = \tau_0, \quad G_{r_2} = g_3(\varphi_{r_2}(s)) (i\tau_1),\\ 
		& \quad \varphi_{r_2}(1)=\varphi_{r_2}(3) = 0, \quad \varphi_{r_2}(2) = \varphi_{r_2}(4) = \theta_{\sigma r_1}.
		\end{align}
		\item
		Class (0,1,1): $\theta_{\sigma r_1}, \theta_{\sigma r_2} = 0,\pi  $, \quad [4 solutions]
		\begin{align}
		\nonumber
		&G_\sigma = g_3(\varphi_\sigma(s)), \\ \nonumber
		&~\varphi_\sigma(1)=\varphi_\sigma(4) = 0, ~ \varphi_\sigma(2)=\varphi_\sigma(3) = \theta_{\sigma r_1},\\ \nonumber
		&G_{r_1} = i\tau_1, \quad G_{r_2} = g_3(\varphi_{r_2}(s)) (i\tau_1), \\ 
		&~\varphi_{r_2}(1)=\varphi_{r_2}(3) = 0, ~ \varphi_{r_2}(2)=\varphi_{r_2}(4) = \theta_{\sigma r_1}+\theta_{\sigma r_2}.
		\end{align}
		\item
		Class (1,1,1): $ \theta_\sigma, \theta_{r_1 r_2} = 0,\pi $, \quad [4 solutions]
		\begin{align}\nonumber
		&G_\sigma = g_3(\varphi_\sigma (s))(i\tau_1), \\ \nonumber
		&\quad \varphi_\sigma(1)=\varphi_\sigma(3) = \theta_\sigma, \quad \varphi_\sigma(2)=\varphi_\sigma(4) = 0,\\ \nonumber
		&G_{r_1}  = (i\tau_1),\\ \label{classH}
		&G_{r_2} = \left\{ \begin{array}{ll}
		g_3\left(\theta_{r_1r_2}/2 \right) (i\tau_1), & s=1,3\\
		g_3\left(-\theta_{r_1r_2}/2\right) (i\tau_1), & s=2,4
		\end{array}
		\right.
		\end{align}
	\end{enumerate}

	\subsection{Time-reversal symmetry}
	Time reversal operation $ {\cal T} $ commutes with all space-group symmetry operations,
	\begin{equation}
	{\cal T}^{-1} U^{-1} {\cal T} U = \boldsymbol{e},
	\end{equation}
	and also with itself
	\begin{equation}
	{\cal T}^2 = \boldsymbol{e}.
	\end{equation}
	Since the gauge groups with translations are all trivial, we have
	\begin{equation}
	G_{\cal T}(n_1,n_2,n_3,s) = g_3(\varphi_{\cal T}(s) + n_1\theta_{{\cal T}1} + n_2\theta_{{\cal T}2} + n_3\theta_{{\cal T}})
	\end{equation}
	or
	\begin{equation}
	G_{\cal T}(n_1,n_2,n_3,s) = g_3(\varphi_{\cal T}(s)+n_1\theta_{{\cal T}1} + n_2\theta_{{\cal T}2} + n_3\theta_{{\cal T}3})(i\tau_1).
	\end{equation}
	The corresponding PSG constraints are
	\begin{align}
	\nonumber
	&G_{\cal T}(i)^2 = g_3(\theta_{\cal T}),\\ \nonumber
	&G_{\cal T}^{-1}(\sigma(i)) G_\sigma^{-1}(i) G_{\cal T}(i) G_{\sigma} (i) = g_3(\theta_{{\cal T}\sigma}),\\ \nonumber
	&G_{\cal T}^{-1}(T_1r_1(i)) G_{r_1}^{-1}(i) G_{\cal T}(i) G_{r_1}(i) = g_3(\theta_{{\cal T}r_1}),\\ \label{gtconstraint}
	&G_{\cal T}^{-1}(T_2r_2(i)) G_{r_2}^{-1}(i) G_{\cal T}(i) G_{r_2}(i) = g_3(\theta_{{\cal T}r_2}).
	\end{align}
	The first equation gives two cases
	\begin{align}\nonumber
	&G_{\cal T}(i) = g_3(\varphi_{\cal T}(s) + n_1\theta_{{\cal T}1} + n_2\theta_{{\cal T}2} + n_3\theta_{{\cal T}3})\\\label{gt1}
	& \qquad \varphi_{\cal T}(s),~ \theta_{{\cal T}1},~ \theta_{{\cal T}2},~ \theta_{{\cal T}3} = 0,\pi; \qquad \theta_{\cal T}=0 \\
	 \nonumber \\ \nonumber
	&G_{\cal T}(i) = g_3(\varphi_{\cal T}(s)+n_1\theta_{{\cal T}1} + n_2\theta_{{\cal T}2} + n_3\theta_{{\cal T}3})(i\tau_1), \\\label{gt2}
	& \qquad \mbox{no constraint}; \qquad \theta_{\cal T}=\pi.
	\end{align}
	In the first case, $ \theta_{\cal T}=0 $ is fixed by the IGG of $ G_{\cal T} $. In the second case, $ \theta_{\cal T}=\pi $ is forced by the constraint and the IGG of $ G_{\cal T} $ is unused.
	\\

	1. For the case of Eq.~(\ref{gt1}), the remaining constraints in (\ref{gtconstraint}) are reduced to
	\begin{align}\nonumber
	&g_3(\theta_{{\cal T}\sigma}) = \eta_{\sigma(s)}\eta_s\\ \nonumber
	&g_3(\theta_{{\cal T}r_1}) =  \eta_{r_1(s)}\eta_{s} \eta_{{\cal T}1}^{n_1+n_2+n_3+A_s+1}\eta_{{\cal T}2}^{-n_3} \eta_{{\cal T}3}^{-n_2} \eta_{{\cal T}1}^{n_1}\eta_{{\cal T}2}^{n_2} \eta_{{\cal T}3}^{n_3},\\
	&g_3(\theta_{{\cal T}r_2}) = \eta_{r_2(s)}\eta_s \eta_{{\cal T}1}^{-n_3}\eta_{{\cal T}2}^{n_1+n_2+n_3+A_s+1} \eta_{{\cal T}3}^{-n_1} \eta_{{\cal T}1}^{n_1} \eta_{{\cal T}2}^{n_2} \eta_{{\cal T}3}^{n_3},
	\end{align}
	where $ \eta_U \equiv g_3(\theta_U) = \pm1 $ due to the constraints $ \varphi_{\cal T}(s)$ and $ \theta_{{\cal T}1,2,3}=0, \pi $. Then,
	\begin{align}\nonumber
	&\eta_{{\cal T}\sigma} = \eta_1\eta_3 = \eta_2\eta_4 = \pm1,\\\nonumber
	&\eta_{{\cal T}1}\eta_{{\cal T}2} \eta_{{\cal T}3} = +1, \\
	&\eta_{{\cal T}1}=\eta_{{\cal T}2}\eta_{{\cal T}3}=+1, \nonumber \\
	\nonumber
	&\eta_{{\cal T}r_1} = \eta_{{\cal T}1}\eta_{{\cal T}3} = \eta_{{\cal T}2}\eta_{{\cal T}4}=\pm1,\\
	&\eta_{{\cal T}2}=+1.
	\end{align}
	In sum, we have 4 solutions, with $ \eta_{{\cal T}\sigma}, \eta_{{\cal T}r_1}=\pm1 $,
	\begin{align}\nonumber
	&G_{\cal T}=\eta_s\tau_0, \\ \label{app:u1result_t1}
	&\quad
	\eta_1=1, ~\eta_2=\eta_{{\cal T}\sigma}\eta_{{\cal T}r_1}, ~\eta_3=\eta_{{\cal T}r_1}, ~\eta_{4}=\eta_{{\cal T}\sigma},
	\end{align}
	Combined with the space group PSG, there are $ 4\times 30=120 $ solutions. However, because of (\ref{trequirement}), the mean-field Hamiltonian becomes sum of two disjoint systems if $G_\mathcal{T}(1) = G_\mathcal{T}(2)$ or $G_\mathcal{T}(2) = G_\mathcal{T}(3)$. Therefore only 30 $U(1)$ projective symmetry groups allow fully connected nearest neighbour mean-field Hamiltonian.
	\\
	
	2. For the case of Eq.~(\ref{gt2}), we solve the remaining constraints separately for the 8 classes Eq.~(\ref{classA})--Eq.~(\ref{classH}). The gauge transformations have the general form
	\begin{align*}
	&G_{\sigma}(s)=\eta_s (i\tau_1)^{n_\sigma}, &\eta&_s=\pm1, ~n_\sigma=0,1,\\
	&G_{r_1} = (i\tau_1)^{n_{r_1}}, &n&_{r_1}=0,1,\\
	&G_{r_2} = \alpha_s(i\tau_3)^{n_{r_2}'} (i\tau_1)^{n_{r_2}}, &\alpha&_{s}=\pm1, ~n_{r_2}, n_{r_2}'=0,1.
	\end{align*}
	The 8 classes are combinations of the following possibilities.
	
	\begin{itemize}
		\item Concerning $ n_\sigma $,
		\begin{multline*}
		g_3(\theta_{{\cal T}\sigma}) = \tau_1 g_3(-\varphi_{\cal T}(\sigma(s))+n_1\theta_{{\cal T}1} + n_2\theta_{{\cal T}2} + n_3\theta_{{\cal T}3}) \tau_1^{n_\sigma}\\
		\cdot g_3(\varphi_{\cal T}(s)+n_1\theta_{{\cal T}1}+n_2\theta_{{\cal T}2}+n_3\theta_{{\cal T}3}).\tau_1\tau_1^{n_\sigma}
		\end{multline*}
		\begin{enumerate}
			\item $ n_\sigma=0 $,
			\begin{align}\nonumber
			\theta_{{\cal T}1}&,~\theta_{{\cal T}2},~ \theta_{{\cal T}3}=0, \pi,\\
			\theta_{{\cal T}\sigma}&=\varphi_{\cal T}(4)-\varphi_{\cal T}(1) \nonumber \\ \label{nsigma}
			&= \varphi_{\cal T}(3)-\varphi_{\cal T}(2)=0,\pi.
			\end{align}
			\item $ n_{\sigma}=1 $,
			\begin{align}
			\theta_{{\cal T}\sigma}&=\varphi_{{\cal T}}(1)+\varphi_{\cal T}(4) \nonumber \\
			&=\varphi_{\cal T}(2)+\varphi_{\cal T}(3).
			\end{align}
		\end{enumerate}
		
		\item Concerning $ n_{r_1} $,
		\begin{multline*}
		g_3(\theta_{{\cal T}r_1}) = \tau_1 g_3(-\varphi_{\cal T}(r_1(s))-(n_1+n_2+n_3+A_s+1)\theta_{{\cal T}1})\\
		\cdot g_3(-n_3\theta_{{\cal T}2}-n_2\theta_{{\cal T}3}) \tau_1^{n_{r_1}} \\
		\cdot g_3(\varphi_{{\cal T}}(s) + n_1\theta_{{\cal T}1}+n_2\theta_{{\cal T}2} + n_3\theta_{{\cal T}3}) \tau_1 \tau_1^{n_{r_1}}.
		\end{multline*}
		\begin{enumerate}
			\item $ n_{r_1}=0 $,
			\begin{align}
			\nonumber
			\theta_{{\cal T}1} &=\theta_{{\cal T}2}-\theta_{{\cal T}3}=0,\pi,\\ \nonumber
			\theta_{{\cal T}r_1} &= \varphi_{\cal T}(3)-\varphi_{\cal T}(1)+\theta_{{\cal T}1} \nonumber \\ \nonumber
			&= \varphi_{\cal T}(4)-\varphi_{\cal T}(2)+\theta_{{\cal T}1}\\ \label{gt2m1case1}
			&= \varphi_{\cal T}(1)-\varphi_{\cal T}(3)
			= \varphi_{\cal T}(2)-\varphi_{\cal T}(4)
			\end{align}
			\item $ n_{r_1}=1 $,
			\begin{align}
			\nonumber
			&\theta_{{\cal T}1}=0, \quad \theta_{{\cal T}2}+\theta_{{\cal T}3}=0,\\
			&\varphi_{{\cal T}}(3)+\varphi_{\cal T}(1) = \varphi_{\cal T}(4)+\varphi_{\cal T}(2)=\theta_{{\cal T}r_1}.
			\end{align}
		\end{enumerate}
		\item Concerning $ n_{r_2} $,
		\begin{multline*}
		g_3(\theta_{{\cal T}r_2}) =\tau_1 g_3(-\varphi_{\cal T}(r_2(s)) -n_3\theta_{{\cal T}1})\\
		\cdot g_3(-(n_1+n_2+n_3+A_s+1)\theta_{{\cal T}2} -n_1\theta_{{\cal T}3})\tau_1^{n_{r_2}} \tau_3^{n_{r_2}'} \\
		\cdot g_3(\varphi_{\cal T}(s)+ n_1\theta_{{\cal T}1} + n_2 \theta_{{\cal T}2} + n_3\theta_{{\cal T}3})\tau_1 \tau_3^{n_{r_2}'} \tau_1^{n_{r_2}'}.
		\end{multline*}
		We see that for $ n_{r_2}'=0, 1 $, the difference is a possible minus sign on the right hand side which amounts to redefine $ \theta_{{\cal T}r_2}\sim \theta_{{\cal T}r_2}+\pi $.
		\begin{enumerate}
			\item $ n_{r_2}=0$,
			\begin{align}
			\nonumber
			\theta_{{\cal T}2} &= \theta_{{\cal T}1}-\theta_{{\cal T}3} = 0,\pi, \\\nonumber
			\theta_{{\cal T}r_2} &= \varphi_{\cal T}(3)-\varphi_{\cal T}(1)+\theta_{{\cal T}2} \nonumber\\ \nonumber
			&=\varphi_{\cal T}(4)-\varphi_{\cal T}(2)+\theta_{{\cal T}2} \\
			&=\varphi_{\cal T}(1)-\varphi_{\cal T}(3) = \varphi_{\cal T}(2)-\varphi_{\cal T}(4).
			\end{align}
			\item $ n_{r_2}=1 $,
			\begin{align}
			\nonumber
			\theta_{{\cal T}2}&=0,\quad \theta_{{\cal T}1}+ \theta_{{\cal T}3}=0,\\ \label{gt2m1case2}
			\theta_{{\cal T}r_2} &= \varphi_{\cal T}(1)+\varphi_{\cal T}(3) = \varphi_{\cal T}(2) + \varphi_{\cal T}(4).
			\end{align}
		\end{enumerate}
	\end{itemize}
	From (\ref{nsigma})--(\ref{gt2m1case2}) we can assemble the results for the 8 classes.
	\begin{itemize}
		\item Class (0,0,0): 4 solutions with $ \eta_{{\cal T}\sigma}, \eta_{{\cal T}r_1}=\pm1 $,
		\begin{align}
		\nonumber
		&G_{\cal T}(s)=\lambda_s i\tau_1,\\ \label{app:u1result_t2_eq1}
		&\lambda_1=1, ~\lambda_2=\eta_{{\cal T}\sigma} \eta_{{\cal T}r_1}, ~\lambda_3=\eta_{{\cal T}r_1}, ~\lambda_4=\eta_{{\cal T}\sigma}.
		\end{align}
		
		\item Class (0,1,0): 2 solutions with $ \eta_{{\cal T}\sigma}=\pm1 $. Note the IGG of $ G_{r_1} $ sets $ \theta_{{\cal T}r_1}=0 $.
		\begin{align}
		&G_{\cal T}(1,3)=i\tau_1, &&
		G_{\cal T}(2,4)=\eta_{{\cal T}\sigma} (i\tau_1).
		\end{align}
		
		\item Class (0,0,1): 2 solutions with $ \eta_{{\cal T}\sigma}=\pm1 $. The IGG of $ G_{r_2} $ sets $ \theta_{{\cal T}r_2}=0 $.
		\begin{align}
		&G_{\cal T}(1,3)=i\tau_1, &&
		G_{\cal T}(2,4)=\eta_{{\cal T}\sigma} (i\tau_1).
		\end{align}
		
		\item Class (1,0,0): 1 solution. The IGG of $ G_\sigma, G_{r_1}, G_{\cal T} $ set $ \theta_{{\cal T}\sigma}, \theta_{{\cal T}r_1}, \varphi_{{\cal T}}(1)=0 $ respectively.
		\begin{equation}
		G_{\cal T}=i\tau_1.
		\end{equation}
		
		\item Class (1,1,0): 2 solutions with $ \eta_{{\cal T}\sigma}=\pm1 $. The IGG of $ G_{\cal T}, G_{r_1} $ set $ \varphi_{\cal T}, \theta_{{\cal T}r_1}=0 $ respectively.
		\begin{align}
		&G_{\cal T}(1,3)=i\tau_1,&& G_{\cal T}(2,4)=\eta_{{\cal T}\sigma}(i\tau_1).
		\end{align}
		
		\item Class (1,0,1): 2 solutions with $ \eta_{{\cal T}\sigma}=\pm1 $. The IGG of $ G_{\cal T}, G_{r_2} $ set $ \varphi_{\cal T}(1), \theta_{{\cal T}r_2}=0 $ respectively.
		\begin{align}
		&G_{\cal T}(1,3)=i\tau_1, &&
		G_{\cal T}(2,4)=\eta_{{\cal T}\sigma}(i\tau_1).
		\end{align}
		
		\item Class (0,1,1): 2 solutions with $ \eta_{{\cal T}\sigma}=\pm1 $. The IGG of $ G_{\cal T}, G_{r_1} $ set $ \varphi_{\cal T}(1), \theta_{{\cal T}r_1}=0 $.
		\begin{align}
		&G_{\cal T}(1,3)=i\tau_1, &&
		G_{\cal T}(2,4)=\eta_{{\cal T}\sigma} (i\tau_1).
		\end{align}
		
		\item Class (1,1,1): 2 solutions with $ \eta_{{\cal T}\sigma}=\pm1 $. The IGG of $ G_{\cal T}, G_{r_1} $ set $ \varphi_{\cal T}, \theta_{{\cal T}r_1}=0 $.
		\begin{align}\label{app:u1result_t2_eq8}
		G_{\cal T}(1,3)= i\tau_1, &&
		G_{\cal T}(2,4) = \eta_{{\cal T}\sigma}(i\tau_1).
		\end{align}
	\end{itemize}
	In sum, combined with the spatial PSG's, there are $ 4\times2+2\times2 + 2\times2 + 8 + (2\times 4) \times 4= 56 $ solutions.
	
	There are total $ 120+56=176 $ U(1) projective extension of the symmetry group. Among those, 30 solutions have the time reversal gauge group $ G_{\cal T}(s=1,3)=-G_{\cal T}(s=2,4)=\tau_0 $. Such a gauge group ensures the sublattice symmetry of the mean-field ans\"{a}ts, and that all of the NN bonds can have non-vanishing amplitudes. Thus, those 30 solutions can all support the topologically protected nodal line Fermi surface, like the ``root" U(1) states of the Kitaev spin liquids shown in the main text.
	
	\subsection{Complete classification of $ Z_2 $ algebraic PSG}\label{app:z2result}
	The calculation for $ Z_2 $ spin liquids on hyperhoneycomb lattices is relatively simple. So we skip the details here and directly present the result. There are total $ 160 $ solutions.
	\begin{enumerate}
		\item When $ G_{\cal T}^2=+1 $, we have six unfixed IGG elements, $ \eta_{{\cal T}\sigma},\eta_{{\cal T}r_1}, \eta_\sigma, \eta_{\sigma r_1}, \eta_{\sigma r_2}, \eta_{r_1 r_2}=\pm1 $. Hence there are total $ 2^6=64 $ solutions.
		
		\begin{align}
		\nonumber
		&G_{\cal T}(s)=\gamma_s\tau_0,\\ \nonumber
		&\qquad \gamma_1=1, \gamma_2=\eta_{{\cal T}r_1}\eta_{{\cal T}\sigma}, \gamma_3=\eta_{{\cal T}r_1}, \gamma_4=\eta_{{\cal T}\sigma}\\\nonumber
		&G_{r_1}=\tau_0, \\ \nonumber
		&G_{\sigma} = \lambda_s\tau_0,\\\nonumber
		&\qquad \lambda_1=\eta_\sigma, \lambda_2=\eta_{\sigma r_1}, \lambda_3=\eta_{\sigma r_1}\eta_\sigma, \lambda_4=1.\\\nonumber
		&\text{(a)}~ \eta_{r_1 r_2} = +1: G_{r_2}(s)=\alpha_s\tau_0,\\ \nonumber
		&\qquad \alpha_1=\alpha_3=1, \quad   \alpha_2=\alpha_4=\eta_{\sigma r_1}\eta_{\sigma r_2}\\ \nonumber
		&\text{(b)}~ \eta_{r_1 r_2} = -1: G_{r_2}(s)=\alpha_s(i\tau_3), \\
		&\qquad \alpha_1=-\alpha_3=1, \alpha_4=-\alpha_2=\eta_{\sigma r_1}\eta_{\sigma r_2}.
		\end{align}
		
		Though there are 64 algebraic $Z_2$ PSG solutions, (\ref{trequirement}) prevents completely connected mean-field Hamiltonian if $G_\mathcal{T}(1) = G_\mathcal{T}(2)$ or $G_\mathcal{T}(2) = G_\mathcal{T}(3)$. So there are 16 fully connected, gauge inequivalent mean-field Hamiltonians.
		
		\item When $ G_{\cal T}^2=-1 $, we have seven unfixed IGG elements, $ \eta_{{\cal T}\sigma}, \eta_{{\cal T}r_1},  \eta_{\mathcal{T} r_2}, \eta_\sigma, \eta_{\sigma r_1}, \eta_{\sigma r_2}, \eta_{r_1 r_2}=\pm1 $. It turns out that $\eta_{r_1 r_2} = -1$ whenever $\eta_{\mathcal{T} r_1} \eta_{\mathcal{T} r_2} = -1$. Hence, there are total $ 2^5\times (2+1)=96 $ solutions:
		\begin{align}
		\nonumber
		&G_{\cal T}(s)=\gamma_s(i\tau_2),\\ \nonumber
		& \qquad \gamma_1=1, ~ \gamma_2=\eta_{{\cal T}r_1}\eta_{{\cal T}\sigma }, ~ \gamma_3=\eta_{{\cal T}r_1}, ~ \gamma_4=\eta_{{\cal T}\sigma},\\ \nonumber
		&G_{r_1}=\tau_0,\\ \nonumber
		& G_\sigma(s)=\lambda_s\tau_0,\\ \nonumber
		&\qquad \lambda_1=\eta_\sigma, ~\lambda_2=\eta_{\sigma r_1},~ \lambda_3=\eta_{\sigma r_1}\eta_\sigma, ~\lambda_4=1\\
		\nonumber
		&\text{(a)}~\eta_{\mathcal{T} r_1} \eta_{\mathcal{T} r_2}=1, ~\eta_{r_1 r_2}=1:  G_{r_2}(s)=\alpha_s\tau_0, \\\nonumber
		&\qquad \alpha_1=\alpha_3=1, ~\alpha_2=\alpha_4=\eta_{\sigma r_1}\eta_{\sigma r_2}\\\nonumber
		&\text{(b)}~\eta_{\mathcal{T} r_1} \eta_{\mathcal{T} r_2}=1, ~\eta_{r_1 r_2}=-1: G_{r_2}(s)=\alpha_s (i\tau_2),\\\nonumber
		&\qquad \alpha_1=-\alpha_3=1,~ \alpha_4=-\alpha_2=\eta_{\sigma r_1}\eta_{\sigma r_2}\\ \nonumber
		&\text{(c)}~\eta_{\mathcal{T} r_1} \eta_{\mathcal{T} r_2}=-1, ~\eta_{r_1 r_2}=-1:  G_{r_2}(s)=\alpha_s (i\tau_3),\\
		&\qquad \alpha_1=-\alpha_3=1, ~\alpha_4=-\alpha_2=\eta_{\sigma r_1}\eta_{\sigma r_2}.
		\end{align}
	\end{enumerate}
	
	\section{``Root" U(1) states of the Kitaev spin liquid and the zero mode condition}
	
	\subsection{PSG constraints for Kitaev spin liquid}\label{app:kitaevPSG}
	From the explicit form of Kitaev PSG, Eqs. (\ref{kitaevpsg1}) and (\ref{kitaevpsg2}), we have the IGG as
	\begin{align}\nonumber
	&\eta_{\cal T}=G_{\cal T}^2(s)=+1 \\\nonumber
	&\eta_{{\cal T}\sigma} = G_{\cal T}^{-1} (\sigma^{-1}(s)) G_\sigma^{-1}(s) G_{\cal T}(s)G_\sigma (s)=-1 \\\nonumber
	&\eta_{{\cal T}r_1} = G_{\cal T}^{-1}(r_1^{-1}(s)) G_{r_1}^{-1}(s) G_{\cal T}(s) G_{r_1}(s)=+1\\\nonumber
	&\eta_{{\cal T}r_2} = G_{{\cal T}}^{-1} (r_2^{-1}(s)) G_{r_2}^{-1}(s) G_{\cal T}(s) G_\sigma(s) = +1 \\\nonumber
	&\eta_\sigma = G_{\sigma}(\sigma(s)) G_\sigma(s) = -1\\\nonumber
	&\eta_{\sigma r_1}= G_{\sigma}(\sigma r_1\sigma(s)) G_{r_1}(r_1\sigma(s)) G_\sigma(\sigma(s)) G_{r_1}(s) = +1\\\nonumber
	&\eta_{\sigma r_2} = G_\sigma(\sigma r_2\sigma(s)) G_{r_2}(r_2\sigma(s)) G_\sigma (\sigma(s)) G_{r_2}(s) =+1\\ \nonumber
	&\eta_{r_1r_2} = G_{r_2}(r_2r_1^{-1}r_2^{-1}(s)) G_{r_1}^{-1}(r_2^{-1}(s)) G_{r_2}^{-1}(s) G_{r_1}(s) = -1\\
	\end{align}
	
	\subsection{Zero mode condition} \label{app:determinant}
	
	The $ Z_2 $ Kitaev spin liquid on hyperhoneycomb lattices possess nodal line Fermi surafaces \cite{Schaffer2015}. Such a structure is preserved in its root U(1) states. To see this, we note that the zero mode condition $ \det{\cal H}_{\boldsymbol{k}} = |\det D_{\boldsymbol{k}}|^2 = 0 $.
	Now we use the LU decomposition
	\begin{eqnarray}
	\left(\begin{array}{cc}
	a & b\\
	c & d
	\end{array}\right) =
	\left(\begin{array}{cc}
	a & 0 \\
	c & I
	\end{array}\right)
	\left(\begin{array}{cc}
	I & a^{-1}b\\
	0 & d-ca^{-1}b
	\end{array}\right),
	\end{eqnarray}
	where $ a,b,c,d $ are matrices. Such a decomposition only relies on the assumption that $ a $ has inverse.  Then the determinant
	\begin{eqnarray}
	\nonumber
	\det \left(\begin{array}{cc}
	a & b\\
	c & d
	\end{array}\right) = \det(a) \det (d-ca^{-1}b).
	\end{eqnarray}
	can be computed easily. For the $ D_{\boldsymbol{k}} $ matrices, the zero mode condition corresponds to
	\begin{eqnarray}\nonumber
	0= \det\left(h_{34}^\dagger - (h_{23}+h_{2_z3}e^{-ik_3}) (h_{12}^\dagger)^{-1} \right.\\ \left. \cdot (h_{41_x}e^{ik_1}+h_{41_y}e^{ik_2})\right).\label{det}
	\end{eqnarray}
	The above condition holds except when $ \det(h_{12})=0 $.
	We use Eq.~(\ref{det}) and Eq.~(\ref{hmatrix})--(\ref{hcase8}) to determine the zero mode conditions in the main text.

\end{document}